\documentclass[epj, nopacs]{svjour}

\usepackage{latexsym}
\usepackage{graphics}
\usepackage{amsmath}
\usepackage{amssymb}
\usepackage{color}
\usepackage{mathtools}
\usepackage{a4wide}
\usepackage[colorlinks=true,a4paper=true,backref=false,linktocpage=true,citecolor=blue,urlcolor=blue,linkcolor=blue,pdfpagemode=UseOutlines]{hyperref}
\usepackage[sort&compress,square,comma,numbers]{natbib}
\usepackage{adjustbox}

\def\babar{\mbox{\slshape B\kern-0.1em{\smaller A}\kern-0.1em
    B\kern-0.1em{\smaller A\kern-0.2em R}}\xspace}
\definecolor{Gray}{gray}{0.925}

\newcommand{\be}{\begin{equation}}
\newcommand{\ee}{\end{equation}}
\newcommand{\bea}{\begin{eqnarray}}
\newcommand{\eea}{\end{eqnarray}}

\begin{document}

\title{Updates on the  determination of $\vert V_{cb} \vert$, $R(D^{*})$ and $\vert V_{ub} \vert/\vert V_{cb} \vert$}
\titlerunning{Updates on the  determination of $\vert V_{cb} \vert$, $R(D^{*})$ and $\vert V_{ub} \vert/\vert V_{cb} \vert$}
\authorrunning{G.Martinelli et al.}

\author{G.~Martinelli\inst{1} \and S.~Simula\inst{2} \and L.~Vittorio\inst{3}
}                     

\institute{Physics Department and INFN Sezione di Roma La Sapienza, Piazzale Aldo Moro 5, 00185 Roma, Italy \and 
Istituto Nazionale di Fisica Nucleare, Sezione di Roma Tre, Via della Vasca Navale 84, I-00146 Rome, Italy \and 
LAPTh, Universit\'e Savoie Mont-Blanc and CNRS, F-74941 Annecy, France}

\date{}

\abstract{We present an updated determination of the values of $\vert V_{cb} \vert$, $R(D^*)$ and $\vert V_{ub} \vert/\vert V_{cb} \vert$ based on the new data on semileptonic $B \to D^* \ell \nu_\ell$ decays by the Belle and Belle-II Collaborations and on the recent theoretical progress in the calculation of the  form factors relevant for semileptonic $B \to D^* \ell \nu_\ell$ and $B_s \to K \ell \nu_\ell$ decays. 
In particular we present results derived by applying either the Dispersive Matrix (DM) method of Refs.\,\cite{DiCarlo:2021dzg, Martinelli:2021frl, Martinelli:2021onb, Martinelli:2021myh, Martinelli:2022tte, Martinelli:2022xir} or the more standard Boyd-Grinstein-Lebed (BGL)\,\cite{Boyd:1997kz} approach to the most recent values of the form factors determined in lattice QCD.  
Using all the available lattice results for the form factors from the DM method we get the theoretical value $R^{\rm th}(D^*) = 0.262 \pm 0.009$ and we extract from a bin-per-bin analysis of the experimental data the value $\vert V_{cb} \vert = (39.92 \pm 0.64) \cdot10^{-3}$. Our result for $R(D^*)$ is consistent with the latest experimental world average $R^{\rm exp}(D^*) = 0.284 \pm 0.012$\,\cite{HFLAV_web23} at the $\simeq 1.5\,\sigma$ level. Our value for $\vert V_{cb} \vert$ is compatible with the latest inclusive determinations $\vert V_{cb} \vert^{\rm incl} = (41.97 \pm 0.48) \cdot 10^{-3}$\,\cite{Finauri:2023kte} and $\vert V_{cb} \vert^{\rm incl} = (41.69\pm 0.63) \cdot 10^{-3}$\,\cite{Bernlochner:2022ucr} within $\simeq 2.6$ and $\simeq 2.0$ standard deviations, respectively.
From a reappraisal of the calculations of $\vert V_{ub} \vert / \vert V_{cb} \vert$, we also obtain $\vert V_{ub} \vert / \vert V_{cb} \vert = 0.087\pm 0.009$ in good agreement with the result $\vert V_{ub} \vert / \vert V_{cb} \vert = 0.0844\pm 0.0056$ from the latest FLAG review\,\cite{FlavourLatticeAveragingGroupFLAG:2021npn}.
}

\onecolumn

\maketitle

\section{Introduction}
\label{sec:introduction}

Motivated by several new experimental and theoretical results, in this work we present a new analysis of several quantities particularly relevant in flavor physics, including the Cabibbo-Kobayashi-Maskawa (CKM) matrix element $\vert V_{cb} \vert$, the $\tau / \mu$ ratio of branching fractions $R(D^*)$ and the CKM ratio $\vert V_{ub} \vert/\vert V_{cb} \vert$. On these quantities, in the last few years, there was an intense activity and debate because of the apparent tension between the inclusive and exclusive values of $\vert V_{cb} \vert$ and the discrepancy between some theoretical predictions and the experiments in the determination of $R(D^*)$, which is an important check of Lepton Flavour Universality. Besides that, the CKM ratio $\vert V_{ub} \vert / \vert V_{cb} \vert$ is a quite important parameter which enters in the global determination of the CKM matrix\,\cite{UTfit:2022hsi}.

The main novelties used in the present analysis are:
\begin{itemize}
\item[1)] The new  experimental results for the semileptonic $B \to D^* \ell \nu_\ell$ decays obtained by the Belle\,\cite{Belle:2023bwv} and Belle-II\,\cite{Belle-II:2023okj} Collaborations;
\item[2)] The final published results for the lattice $B \to D^* \ell \nu_\ell$ form factors (FFs) by the FNAL/MILC Collaboration\,\cite{FermilabLattice:2021cdg};
\item[3)] The new results for the $B \to D^* \ell \nu_\ell$  FFs by the HPQCD Collaboration\,\cite{Harrison:2023dzh};
\item [4)]The new results for the $B \to D^* \ell \nu_\ell$ FFs by the JLQCD Collaboration\,\cite{Aoki:2023qpa};
\item[5)] A  new lattice computation of the FFs relevant  for $B_s \to K  \ell \nu_\ell$ decays by the RBC/UKQCD Collaboration\,\cite{Flynn:2023nhi}.
\end{itemize}

We mention that there are other novelties available in the literature, namely a new determination of the FFs relevant for $B \to \pi \ell \nu_\ell$ decays by the JLQCD Collaboration\,\cite{Aoki:2023qpa} and for $B_s\to D_s^* \ell \nu_\ell$ decays in Ref.\,\cite{Harrison:2023dzh}. The analysis of these theoretical data is beyond the aim of the present work and it will be carried out in a separate work.

By using as inputs the new experimental results from 1) and the previous ones published in Ref.\,\cite{Belle:2018ezy} by Belle Collaboration, already used in FLAG '21\,\cite{FlavourLatticeAveragingGroupFLAG:2021npn}, the FFs from 2)-4) and adopting the DM method of Refs.\,\cite{DiCarlo:2021dzg, Martinelli:2021frl, Martinelli:2021onb, Martinelli:2021myh, Martinelli:2022tte, Martinelli:2022xir} we get    
\be 
    \vert V_{cb} \vert = (39.92 \pm 0.64) \cdot 10^{-3}
    \label{eq:vcbex}
\ee
and the theoretical prediction 
\be 
    R(D^*) = 0.262 \pm 0.009 ~ . ~
    \label{eq:RDstar_DM}
\ee 
  
The result in Eq.\,(\ref{eq:vcbex}) is compatible respectively at the $\simeq 2.6\,\sigma$ and $\simeq 2.0\,\sigma$ level with the most recent inclusive determinations $\vert V_{cb} \vert^{\rm{incl}} = (41.97 \pm 0.48) \cdot 10^{-3}$~\cite{Finauri:2023kte} and $\vert V_{cb} \vert^{\rm{incl}} = (41.69 \pm 0.63) \cdot 10^{-3}$~\cite{Bernlochner:2022ucr}. 
Since both the exclusive and the inclusive determinations of $\vert V_{cb} \vert$ are reaching the percent level of accuracy, a first-principles estimate of QED effects in $b \to c$ decays is becoming timely (see also Ref.\,\cite{Bigi:2023cbv}).
Note that using weak processes other than inclusive and exclusive semileptonic $b \to c$ decays, an indication of a large value of $\vert V_{cb} \vert$ was already claimed by the UT\emph{fit} Collaboration in Ref.\,\cite{Alpigiani:2017lpj,UTfit:2022hsi} and more recently by Ref.\,\cite{Buras:2021nns}.
 
The DM method allows to predict the $\tau / \mu$ ratio $R(D^*)$ from theory, obtaining the result\,(\ref{eq:RDstar_DM}), which is compatible with the latest experimental world average $R(D^*) = 0.284 \pm 0.012$\,\cite{HFLAV_web23} at the $\simeq 1.5\,\sigma$ level.

 The CKM ratio $\vert V_{ub} \vert / \vert V_{cb} \vert$ is determined using the LHCb measurement of ${\cal{BR}}(B_s \to K \ell \nu) /$ ${\cal{BR}}(B_s \to D_s \ell \nu)$. The novelty is that a new lattice computation of the FFs from 5) exists for $B_s \to K$ semileptonic decays, which supersedes the results given in Ref.\,\cite{Flynn:2015mha}.  On the other hand, no novelty for the FFs entering the $B_s \to D_s$ semileptonic  decays appeared yet. From a  reappraisal of the calculation of $\vert V_{ub} \vert / \vert V_{cb} \vert$, we also obtain $\vert V_{ub} \vert / \vert V_{cb} \vert = 0.087 \pm 0.009$. Our result is properly based on values of the FFs computed only in the $q^2$-region covered by direct lattice results and it is in good agreement with the result $\vert V_{ub} \vert / \vert V_{cb} \vert = 0.0844\pm 0.0056$ from the latest FLAG review\,\cite{FlavourLatticeAveragingGroupFLAG:2021npn}.

This work is divided mainly into two Sections. Section\,\ref{sec:vcb} is devoted to a discussion of the FFs relevant for semileptonic $B \to D^*$ decays, to the derivation of the exclusive value of $\vert V_{cb} \vert$ and to the calculation of $R(D^*)$. Section\,\ref{sec:vubsvcb} contains a comparison among different lattice calculations of the FFs relevant for $B_s \to K$ semileptonic decays and a new determination of $\vert V_{ub} \vert / \vert V_{cb} \vert$ from the ratio ${\cal{BR}}(B_s \to K \ell \nu_\ell) / {\cal{BR}}(B_s \to D_s \ell \nu_\ell)$ ratio. 
Our final considerations and outlooks can be found in Section\,\ref{sec:conclusions}.

\section{$\vert V_{cb} \vert$ using lattice form factors}
\label{sec:vcb}

This section is divided into three parts. In the first one we will discuss and compare the FFs as a function of the four-momentum transfer obtained on the lattice by various collaborations\,\cite{FermilabLattice:2021cdg, Harrison:2023dzh, Aoki:2023qpa} and their extension to the full allowed kinematical range using either the DM method or the more standard BGL\,\cite{Boyd:1997kz} approach. 
In the second part we discuss and compare the values of $\vert V_{cb}\vert$ extracted from the experimental data by means of a bin-per-bin analysis and, on the basis of the present information, we derive our best final estimate for this quantity.
In the third part we present the result of our analysis for the $\tau / \mu$ ratio $R(D^*)$.

\subsection{The unitary bands for the form factors}
\label{sec:FFs}

\begin{figure}[htb!]
\begin{center}
\includegraphics[scale=0.250]{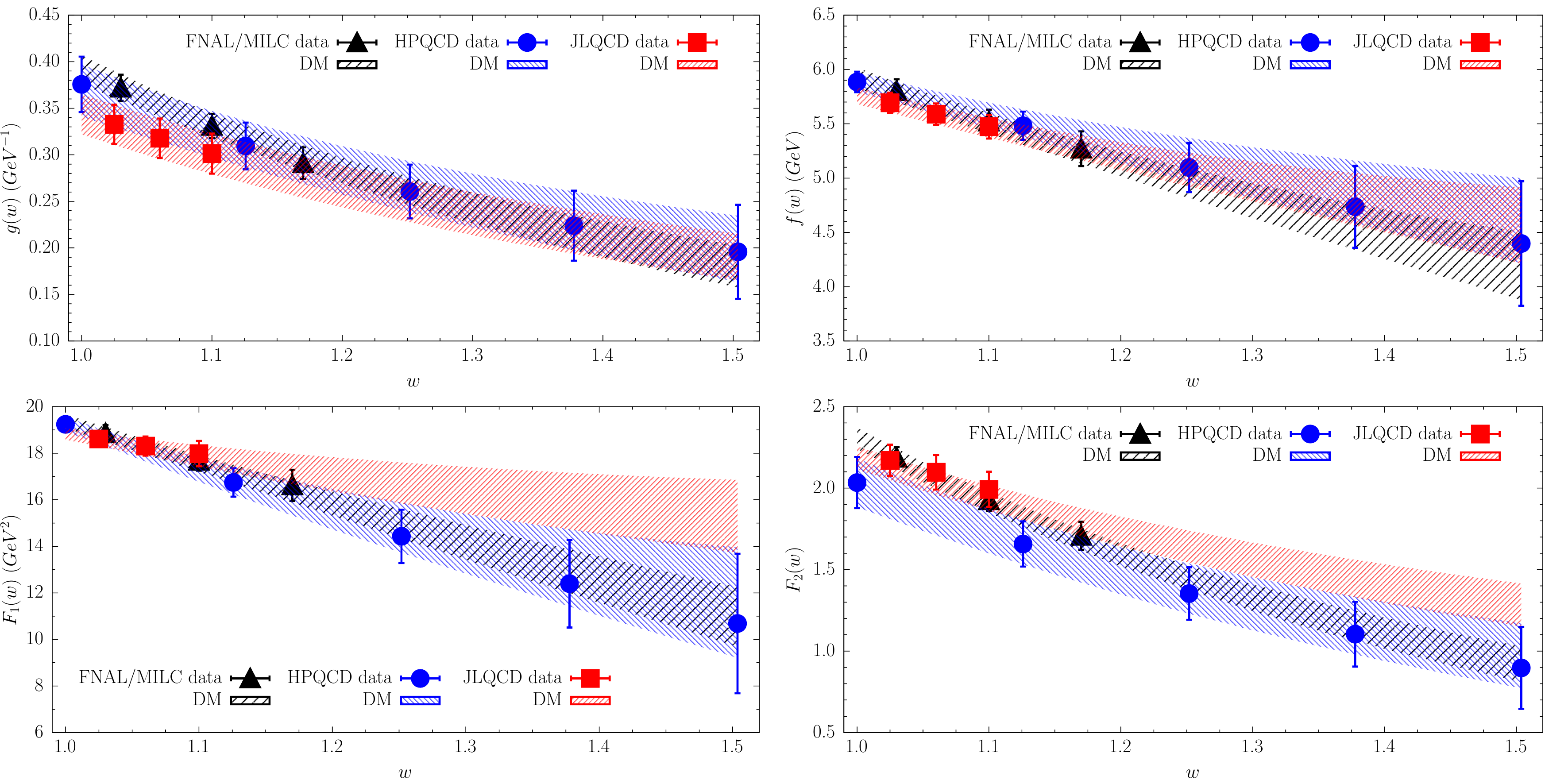}
\caption{\it \small The FFs $g(w)$, $f(w)$, $F_1(w)$ and $F_2(w)$ of the semileptonic $B \to D^* \ell \nu_\ell$ decays computed by the FNAL/MILC (triangles), HPQCD (circles) and JLQCD (squares) Collaborations versus the recoil variable $w$. The bands (at $1\,\sigma$ level) obtained by applying the DM method separately to each of the three lattice data sets are shown after imposing the KCs\,(\ref{eq:KCs}).}
\label{fig:FFsDM}
\end{center}
\end{figure}

In Fig.\,\ref{fig:FFsDM} we show the relevant FFs from the three independent lattice calculations\,\cite{FermilabLattice:2021cdg, Harrison:2023dzh, Aoki:2023qpa} as a function of the recoil variable $w = (m_B^2 + m_{D^*}^2 - q^2) / (2\,m_Bm_{D^*})$.  
Note that only in Ref.\,\cite{Harrison:2023dzh} the lattice FFs are computed on (almost) the full accessible kinematical range ($1 \leq w \leq w_{\rm max} \simeq 1.50$), whereas for the other collaborations the maximum value at which the FFs have been computed is smaller than $w \simeq 1.2$.  
Thus, in order to compare the results of the different collaborations over the full kinematical range, which is also relevant for the extraction of $\vert V_{cb} \vert$ from the experimental data and for the calculation of $R(D^*)$, we have to extrapolate the results of Refs.\,\cite{FermilabLattice:2021cdg} and \cite{Aoki:2023qpa} up to $w_{max}$,  which corresponds to the maximum value of the spatial momentum  of the $D^*$ meson in the $B$-meson rest frame. It is precisely the large-$w$ region where potentially dangerous discretisation effects may play a role. 
 
In order to extrapolate the FFs in $w$,  we apply separately the DM method to the  lattice computations of the different collaborations, taking into account 
the following kinematical constraints (KCs):
\be
    \label{eq:KCs}
    F_1(1)  =  m_B (1 - r) f(1) \, ,  \qquad 
    F_2(w_{max}) = \frac{2}{1 - r^2} ~ \frac{F_1(w_{max})}{m_B^2} ~
\ee
with  $r \equiv m_{D^*}/m_B \simeq 0.38$.
A detailed description of the DM approach, including the precise definitions of the four FFs and the implementation of the KCs\,(\ref{eq:KCs}) as well as of the unitary constraints that must be satisfied by the FFs $g$, $(f,  F_1)$ and $F_2$, can be found in Refs.~\cite{DiCarlo:2021dzg, Martinelli:2021onb, Martinelli:2021myh, Martinelli:2022xir} (see also Appendix\,\ref{sec:DM}). As for the unitary bounds we make use of the non-perturbative susceptibilities evaluated on the lattice in Ref.\,\cite{Martinelli:2021frl} (see also Appendix\,\ref{sec:sampling}).

From Fig.\,\ref{fig:FFsDM} we observe that: 
\begin{itemize}
\item[i)]  there are some differences between the values of $F_2(w)$ from the updated HPQCD determinations of Ref.\,\cite{Harrison:2023dzh} and those of the other two collaborations even in the region where all the three groups present results. These differences, however, have a minor impact on the extraction of $\vert V_{cb}\vert$ from the experimental data for light leptons, as it is the case of the Belle\,\cite{Belle:2018ezy, Belle:2023bwv} and  Belle-II\,\cite{Belle-II:2023okj} data. Nevertheless, it is important to extrapolate also the FF $F_2$ in $w$ because of the KC at $w = w_{max}$ (see Refs.\,\cite{Martinelli:2021onb, Martinelli:2021myh, Martinelli:2022xir}).
\item[ii)] although at $w\sim 1$ the values of $F_2(w)$ from FNAL/MILC and JLQCD are close, the bands of the extrapolated values at large $w$ are quite  different;  
\item[iii)] the results for $g(w)$, $f(w)$ and $F_1(w)$ are in reasonable agreement in the range of $w$ where all the three collaborations have computed  the FFs (i.e.~$w \leq 1.2$);
\item[iv)] the allowed band of the extrapolated values of $F_1(w)$ at large $w$ using the inputs from JLQCD\,\cite{Aoki:2023qpa}, however, is different from the bands obtained for this quantity using either the FNAL/MILC\,\cite{FermilabLattice:2021cdg} or the HPQCD inputs\,\cite{Harrison:2023dzh}. This difference originates from the rather different slope of $F_1(w)$ at small values of $w$.  In the next subsection we will see that this difference has rather strong consequences on the extraction of $\vert V_{cb} \vert$ and on the calculation of $R(D^*)$;
\item[v)] the FNAL/MILC and JLQCD FFs are based on the use of $N_f = 2+1$ gauge configurations, while only the HPQCD FFs take into account the effects of a dynamical charm quark in the sea. Generally speaking, these effects are expected to be subdominant with respect to the present uncertainties of the FFs. Note also that the differences observed among the FNAL/MILC and JLQCD FFs cannot be ascribed to the absence of a dynamical charm quark in the sea. The latter one might have a role in the case of the differences observed in the case of $F_2$, which (as already stressed) has a minor impact on the extraction of $\vert V_{cb}\vert$ from the experimental data for light leptons.
\end{itemize}

\begin{figure}[htb!]
\begin{center}
\includegraphics[scale=0.250]{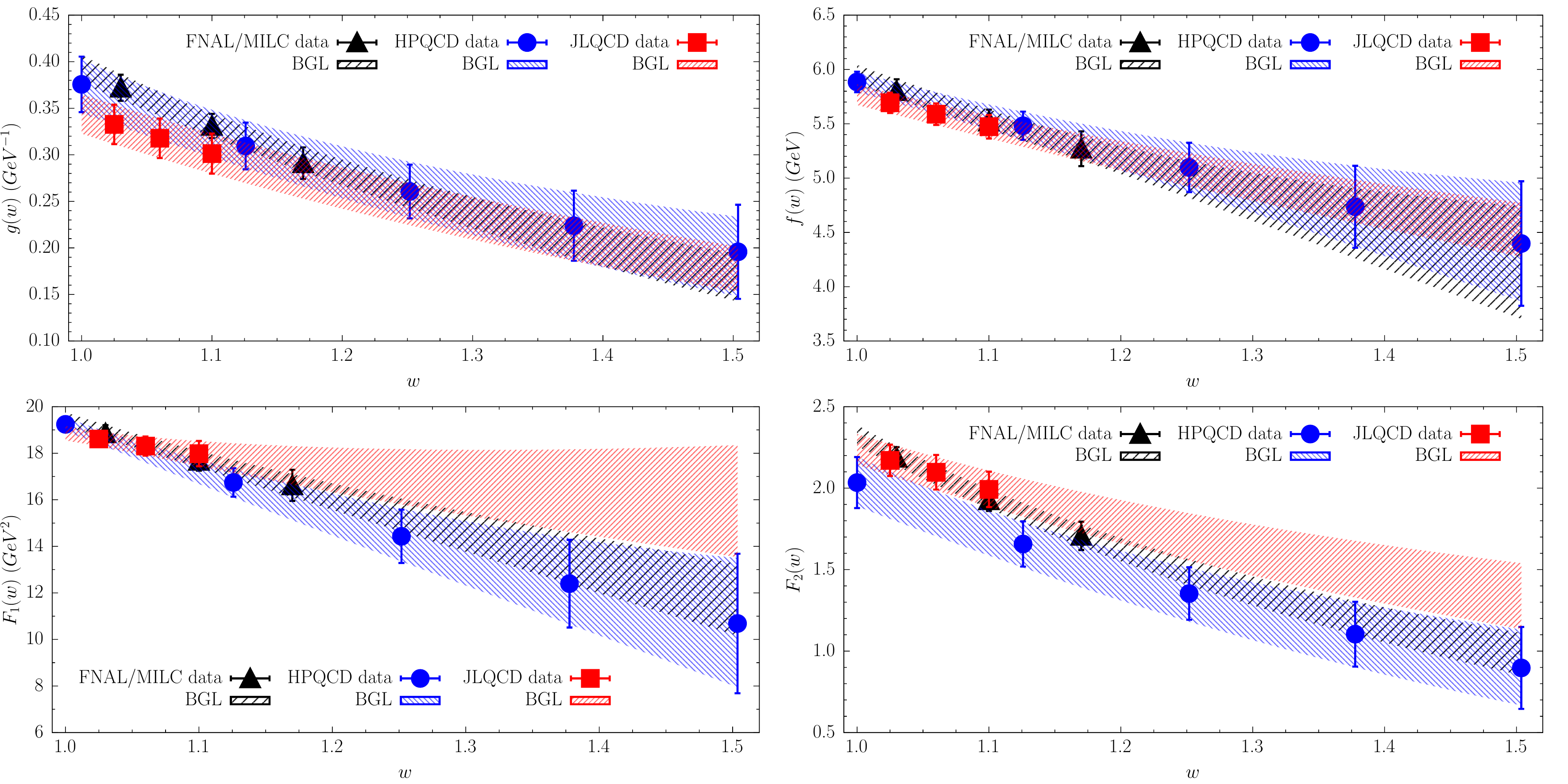}
\caption{\it \small The same as in Fig.\,\ref{fig:FFsDM}, but  the bands are derived using  the BGL method\,\cite{Boyd:1997kz} after imposing both the unitary filters and the KCs\,(\ref{eq:KCs}) according to the procedure proposed in Ref.\,\cite{Simula:2023ujs}. The BGL $z$-expansions are truncated after the quadratic term for the FFs $g$ and $f$, and after the cubic term for the FFs $F_1$ and $F_2$.}
\label{fig:FFsBGL}
\end{center}
\end{figure}

In order to demonstrate that the difference between JLQCD and the other collaborations is not an artefact of the DM approach, we show in Fig.\,\ref{fig:FFsBGL}  the extrapolations of the FFs using the BGL method\,\cite{Boyd:1997kz} complemented by both the unitary and kinematical constraints. To guarantee that the BGL fitting procedure satisfies exactly unitarity, we follow the procedure described in Section VIII of Ref.\,\cite{Simula:2023ujs}.
We observe that  the discrepancy at large values of $w$ between JLQCD and the other two collaborations persists also in the BGL approach. 
Note that  the DM method produces narrower bands for the extrapolation of the FFs. This  is a consequence of the automatic elimination of the subset of input values of the FFs which, although allowed by the uncertainties of the lattice calculations, do not satisfy unitarity (and the KCs). See Ref.\,\cite{Simula:2023ujs} for a detailed discussion on the features of the DM unitary filter. 

Before closing this Section we stress that the FFs predicted by the DM method, as those shown in Fig.\,\ref{fig:FFsDM}, can be easily and effectively incorporated in the calculation of observables of interest for investigations within the Standard Model and beyond, as it has been recently done in Ref.\,\cite{Fedele:2023ewe}, using the HEPfit package\,\cite{DeBlas:2019ehy}, and in Ref.\,\cite{Guadagnoli:2023ddc}, using the Flavio package\,\cite{Straub:2018kue}.

\subsection{Determination of $\vert V_{cb} \vert$}
\label{sec:Vcb}

The large differences among the FFs computed  by the three collaborations, and in particular for $F_1(w)$ at larger values of $w$, induce substantial differences in the extraction of $\vert V_{cb} \vert$ and in the calculation of $R(D^*)$. Thus, there are two possibilities: 
\begin{itemize}
\item to perform the phenomenological analysis separately for the three collaborations and then combine the final results taking into account the correlations due to the use of the same experimental data, as it will be described in Section\,\ref{sec:trino};
\item to consider all the results for the FFs and perform a single phenomenological analysis\footnote{When combining the lattice calculations of the three collaborations the FFs can be considered uncorrelated, since their determination is based on different gauge configurations.}, as carried out in Section\,\ref{sec:unos}.
\end{itemize}

 The large differences/tensions eventually present also in the experimental data, however, require to select a procedure to combine the results. Our reference choice is the evaluation of correlated averages (see, e.g., later on Eqs.\,(\ref{eq:muVcb})-(\ref{eq:sigmaVcb})). This procedure requires a precise determination of the correlations among different calculations and it may suffer from rather disturbing effects, like the well-known D'Agostini bias\,\cite{DAgostini:2020pim} or abnormally high values of the $\chi^2$ variable. Such effects may occur typically in the case of large  positive values of the correlations. In order to check whether correlated averages are trustable for our calculations, we have employed the same strategy adopted in Ref.\,\cite{Martinelli:2021myh}. Such  alternative procedure combines the mean values of the different calculations, $x_k$, and uncertainties, $\sigma_k$, through the formul\ae (see Ref.\,\cite{EuropeanTwistedMass:2014osg})
\bea
    \label{eq:average28}
    \mu_x & = & \sum_{k=1}^N w_k x_k ~ , ~ \\[2mm]
    \label{eq:sigma28}
    \,\sigma_x^2 & = & \sum_{k=1}^N \,w_k \sigma_k^2 + \sum_{k=1}^N \, w_k (x_k - \mu_x)^2 ~ , ~
\eea
where the second term in the r.h.s.~of Eq.\,(\ref{eq:sigma28}) accounts for the spread of the values due to systematic errors in the  calculations/experimental  measurements, while $w_k$ represents the weight assigned to the $k$-th calculation $x_k$  and satisfies the normalization condition $\sum_{k=1}^N w_k = 1$. In what follows we consider two choices of the weight $w_k$: the first one is the democratic choice  $w_k = 1 / N$ and the second one is $w_k \propto  \sigma_k^{-2}$. 

Typically, Eq.\,(\ref{eq:sigma28}) tends to give larger errors with respect to the one of other approaches, like the correlated average procedure.  Our aim is to compare the mean values obtained by the correlated average procedure and by Eq.\,(\ref{eq:average28}). If the two agrees, then we can trust the result obtained by the correlated average procedure and adopt also its more precise error.
 
Other possible averaging procedure are the one of the Particle Data Group (PDG) \cite{ParticleDataGroup:2022pth} or the one suggested in Refs.\,\cite{DAgostini:1999niu, DAgostini:2020vsk}. We have checked that for the quantities of interest in this work all the above procedures give similar mean values well compatible within the uncertainties.

\subsubsection{$\vert V_{cb} \vert$ from three independent analyses}
\label{sec:trino}

Our analysis is based on three sets of measurements of the differential decay widths performed by the Belle\,\cite{Belle:2023bwv, Belle:2018ezy} and Belle-II\,\cite{Belle-II:2023okj} Collaborations for semileptonic $B \to D^* \ell \nu_\ell$ decays in terms of four different kinematical variables\footnote{We do not make use of the results of the BaBar Collaboration given in Ref.\,\cite{BaBar:2019vpl}, since the final synthetic data are based on parameterization-dependent fits of the FFs.}, namely $x = w, \cos \theta_l, \cos \theta_v, \chi$ (see Ref.\,\cite{Martinelli:2021onb} for the expression of the four-dimensional differential decay widths and Refs.\,\cite{Belle:2023bwv, Belle:2018ezy, Belle-II:2023okj} for the specific values of the four variables $x$ in each bin).

As first proposed by the present authors in Refs.\,\cite{Martinelli:2021onb, Martinelli:2021myh}, we may determine values of $\vert V_{cb} \vert$ by performing a bin-per-bin study of the experimental data. Indeed, the comparison between the theoretical differential rates integrated over specific bins, and the corresponding experimental measurements, gives determinations of $\vert V_{cb} \vert$ in which the shape of the theoretical FFs is not influenced by the shape of the experimental differential rates. On the contrary, the procedure of fitting the FFs using also experimental data, the so-called {\em joint fit} adopted, e.g., in Ref.\,\cite{FermilabLattice:2021cdg}, may generate weird results in the study of $B \to D^* \ell \nu_\ell$ decays, when it is applied without caution to experimental differential rates having slopes different from the theoretical  predictions (see Ref.\,\cite{DM_web22}). Instead, the bin-per-bin analysis provides detailed information about the agreement/disagreement of the shapes of the distributions between theory and experiments in the different phase space regions and for different kinematical variables.
Our procedure is  the following. 

\begin{itemize}

\item The Belle experimental data~\cite{Belle:2023bwv, Belle:2018ezy} are given in the form of 10-bins distribution of the quantity $d\Gamma / dx$, where $x$ is one of the four kinematical variables of interest ($x = w, \cos \theta_l, \cos \theta_v, \chi$). The Belle-II data\,\cite{Belle-II:2023okj} are given in the same 10 Belle  bins for the variables $x = \{w,  \cos \theta_v, \chi \}$, while in the case of $x = \cos \theta_l$ the Belle-II bins are only 8\footnote{As for the variable $x = \cos \theta_l$, the Belle-II bins $2 - 8$ correspond to the Belle bins $4 - 10$, while the first bin of Belle-II corresponds to the sum of the first three Belle bins.}. For each kinematical variable $x$ the sum over the bins cover the full kinematical range and, therefore, it must be independent on the choice of $x$. In what follows, for each set of experimental data we simply denote by $\Gamma$ the total rate, by $N_x$ the number of bins of the kinematical variable $x$ and by $N = \sum_x N_x$ the total number of bins (i.e., 40 for each of the two sets of Belle data and 38 for the Belle-II set). For each experiment we consider the available results of the differential decay widths $d\Gamma^{exp} / dx$ for all the bins and the corresponding experimental covariance matrix $\mathbf{C}_{ij}^{exp}$ ($i, j = 1, ..., N$). 

\item We compute the theoretical predictions $d\Gamma^{th} / dx / |V_{cb}|^2$ by generating a set of values of the FFs $g$, $f$, $F_1$ and $F_2$ for each experimental bin. We compute also the corresponding theoretical covariance matrix $\mathbf{C}_{ij}^{th}$ for all the experimental bins ($i, j = 1, ..., N$). 

\item Using multivariate Gaussian distributions and taking into account that experimental data and theoretical predictions are uncorrelated, we generate a sample of values of both $d\Gamma^{exp} / dx$ and $d\Gamma^{th} / dx / |V_{cb}|^2$ for all the bins of each experiment.
Finally, for each event of the sample we compute $|V_{cb}|$ as the square root of the ratio of the experimental over the theoretical differential decay widths for each bin. The distribution of the values of $|V_{cb}|$ turns out to be very well approximated by a Gaussian distribution on which  we compute the mean values $\vert V_{cb} \vert_i$  and the corresponding covariance matrix $\mathbf{C}_{ij}$ among all the experimental bins.

\end{itemize}

As for the experimental covariance matrices we proceed as follows. In the case of the Belle-II data set we read off the covariance matrix directly from the Tables provided in Ref.\,\cite{Belle-II:2023okj}.
In the case of  the Belle data of Ref.\,\cite{Belle:2018ezy}, because of problems with the experimental correlation matrix, we adopt the strategy described in detail in Ref.\,\cite{Martinelli:2021onb}. We consider the relative differential decay rates given by the ratios $(d\Gamma / dx) / \Gamma$ for each variable $x$ and for each bin by using the experimental data.  
In this way we guarantee that the sum over the bins is exactly independent (event by event) of the choice of the variable $x$. 
Hence, we compute a \emph{new} correlation matrix using the events for the ratios $(d\Gamma / dx) / \Gamma$. 
The new correlation matrix has four eigenvalues equal to zero, because the sum over the bins of each of the four variable $x$ is always equal to unity. In other words, the number of independent bins for the Belle ratios is 36 and not 40. 
Then, following Ref.\,\cite{Martinelli:2021onb} a \emph{new} covariance matrix of the experimental data is constructed by multiplying the new correlation matrix by the original uncertainties associated to the measurements.
We then apply our procedure for the extraction of $\vert V_{cb} \vert$ using the new experimental covariance matrix. 
  
 In the case of Ref.\,\cite{Belle:2023bwv} it is not necessary to follow the above procedure, since the experimental data are already given as ratios $(d\Gamma / dx) / \Gamma$ for each bin.
In this  case, however, in order to compare the values of $d\Gamma / dx$ from the three sets of measurements, we have to multiply the experimental value of $(d\Gamma / dx) / \Gamma$ of Ref.\,\cite{Belle:2023bwv} by the total width $\Gamma$ and to construct the experimental covariance matrix according to the procedure described in Ref.\,\cite{Martinelli:2022xir}. 
Following Ref.\,\cite{Belle:2023bwv}, we use the experimental  branching ratios ${\cal{BR}}(B^- \to D^{*0} \ell \nu_\ell) = (5.58 \pm 0.22) \cdot 10^{-2}$ and ${\cal{BR}}(\bar{B}^0 \to D^{*+} \ell \nu_\ell) = (4.97 \pm 0.12) \cdot 10^{-2}$ from HFLAV\,\cite{HeavyFlavorAveragingGroup:2022wzx} as well as the mean lifetimes $\tau_{\bar{B}^0} = 1.520$ ps and $\tau_{B^-} = 1.638$ ps, obtaining for the isospin-averaged total width (according to Eqs.\,(\ref{eq:average28})-(\ref{eq:sigma28}) with $w_k = 1 / N$  and $N = 2$) the value
\be  
    \Gamma(B \to D^* \ell \nu_\ell) = (2.20  \pm 0.09) \cdot 10^{-14} ~ \mbox{GeV} ~ . ~  
 \ee

The presence of the result from the Belle experiment\,\cite{Belle:2018ezy} in the HFLAV average\,\cite{HeavyFlavorAveragingGroup:2022wzx} for the the total branching ratio of the decay channel ${\cal{BR}}(\bar{B}^0 \to D^{*+} \ell \nu_\ell)$ may introduce a correlation between the extracted values of $|V_{cb}|$ from the two Belle experiments. From the weights of the average evaluated in Ref.\,\cite{HeavyFlavorAveragingGroup:2022wzx} we estimate a correlation coefficient equal to $\simeq 0.2$.

\begin{figure}[htb!]
\begin{center}
\includegraphics[scale=0.50]{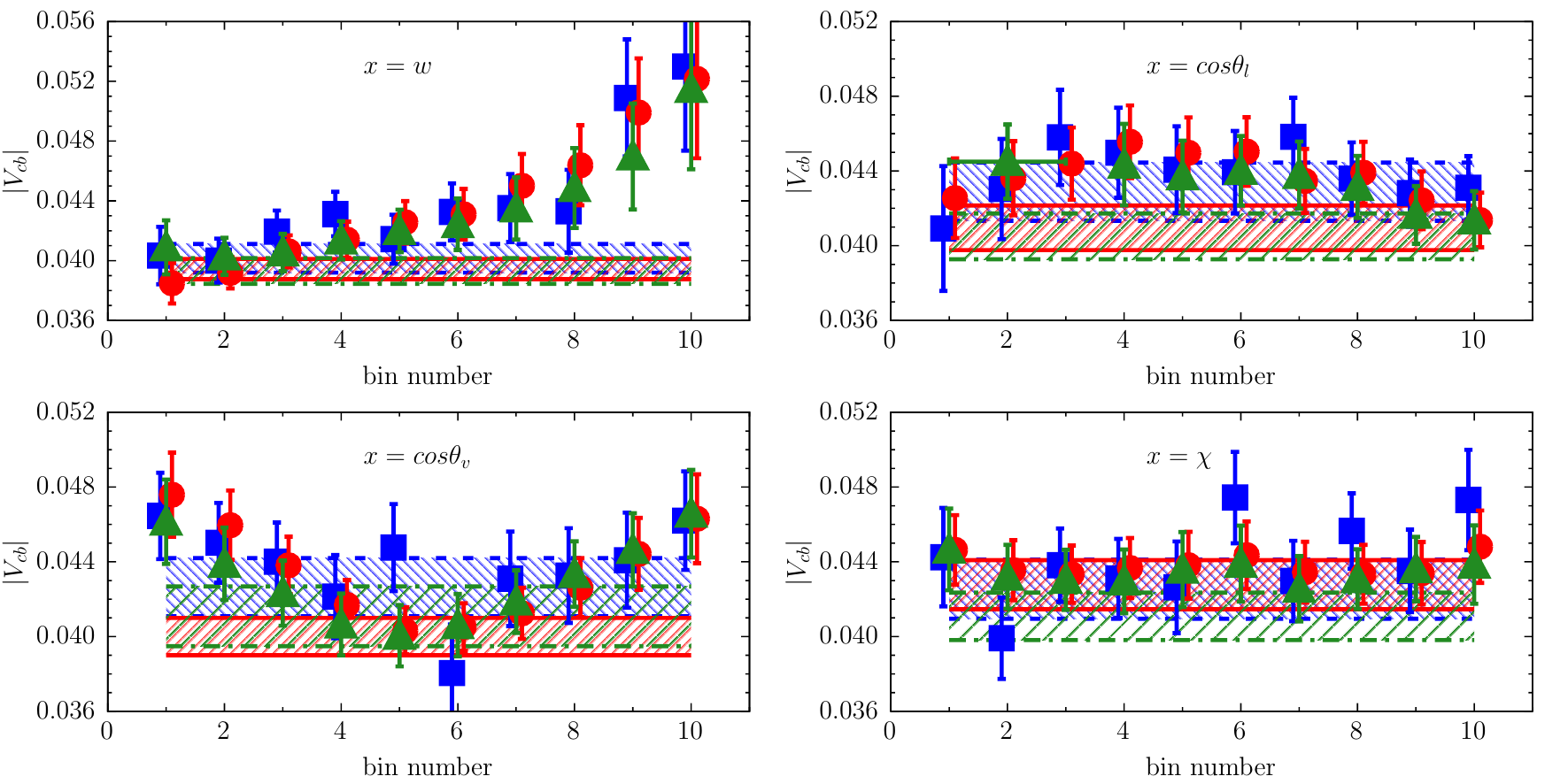}
\caption{\it \small The bin-per-bin estimates of $\vert V_{cb} \vert$ for each kinematical variable $x$ using the experimental data from Refs.\,\cite{Belle:2023bwv} (blue squares),\,\cite{Belle-II:2023okj} (green triangles) and \cite{Belle:2018ezy} (red circles), adopting the extrapolated FFs obtained by the DM method from FNAL/MILC\,\cite{FermilabLattice:2021cdg}. The horizontal blue, green and red bands correspond to the correlated averages evaluated according to Eqs.\,(\ref{eq:muVcb})-(\ref{eq:sigmaVcb}) over the bins.}
\label{fig:Vcb_origFNAL}
\end{center}
\end{figure}

\begin{figure}[htb!]
\begin{center}
\includegraphics[scale=0.50]{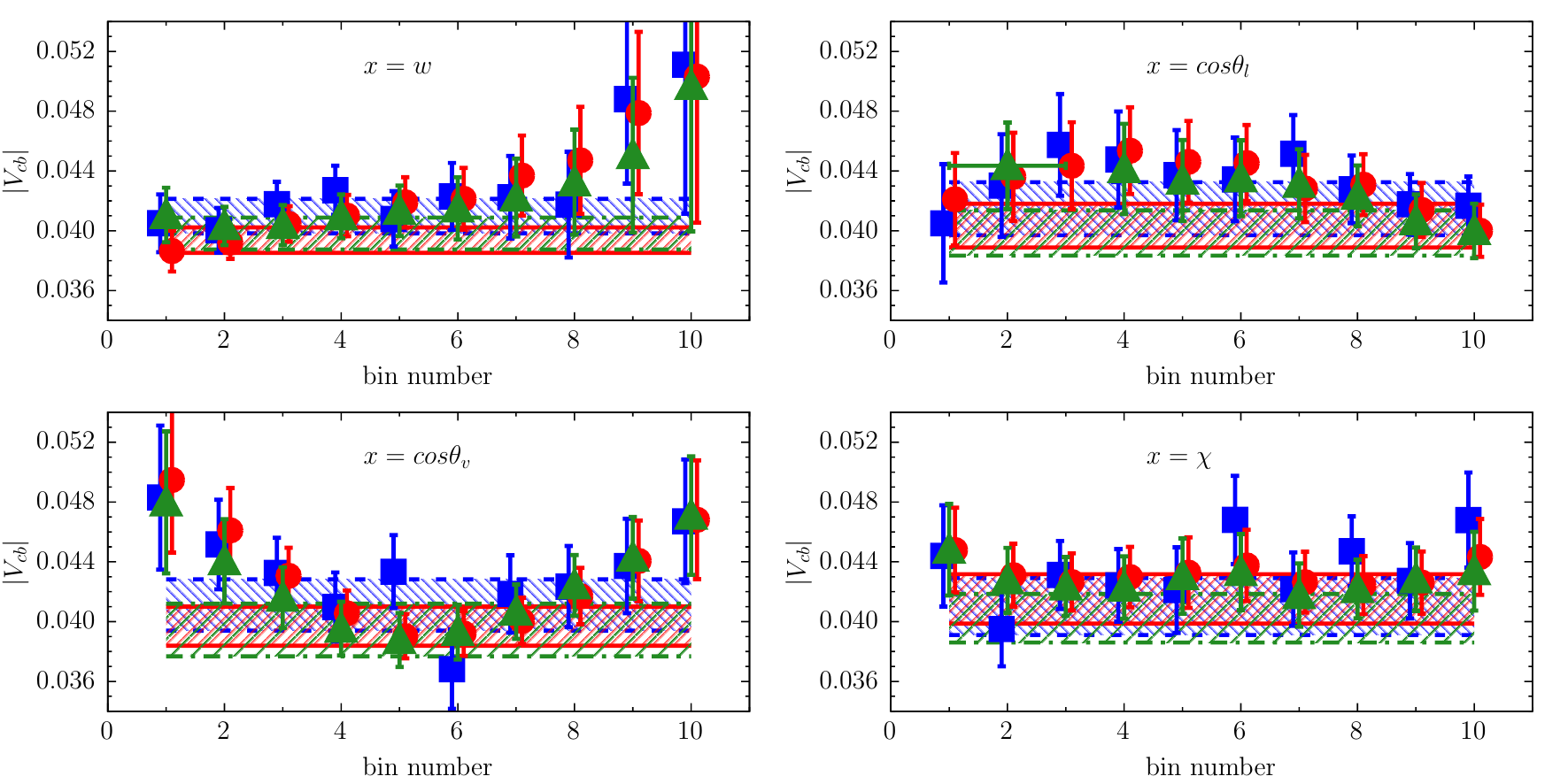}
\caption{\it \small The same as in Fig.\,\ref{fig:Vcb_origFNAL}, but using as inputs for the DM method  the lattice FFs evaluated by the HPQCD Collaboration\,\cite{Harrison:2023dzh}.}
\label{fig:Vcb_origHPQCD}
\end{center}
\end{figure}

\begin{figure}[htb!]
\begin{center}
\includegraphics[scale=0.50]{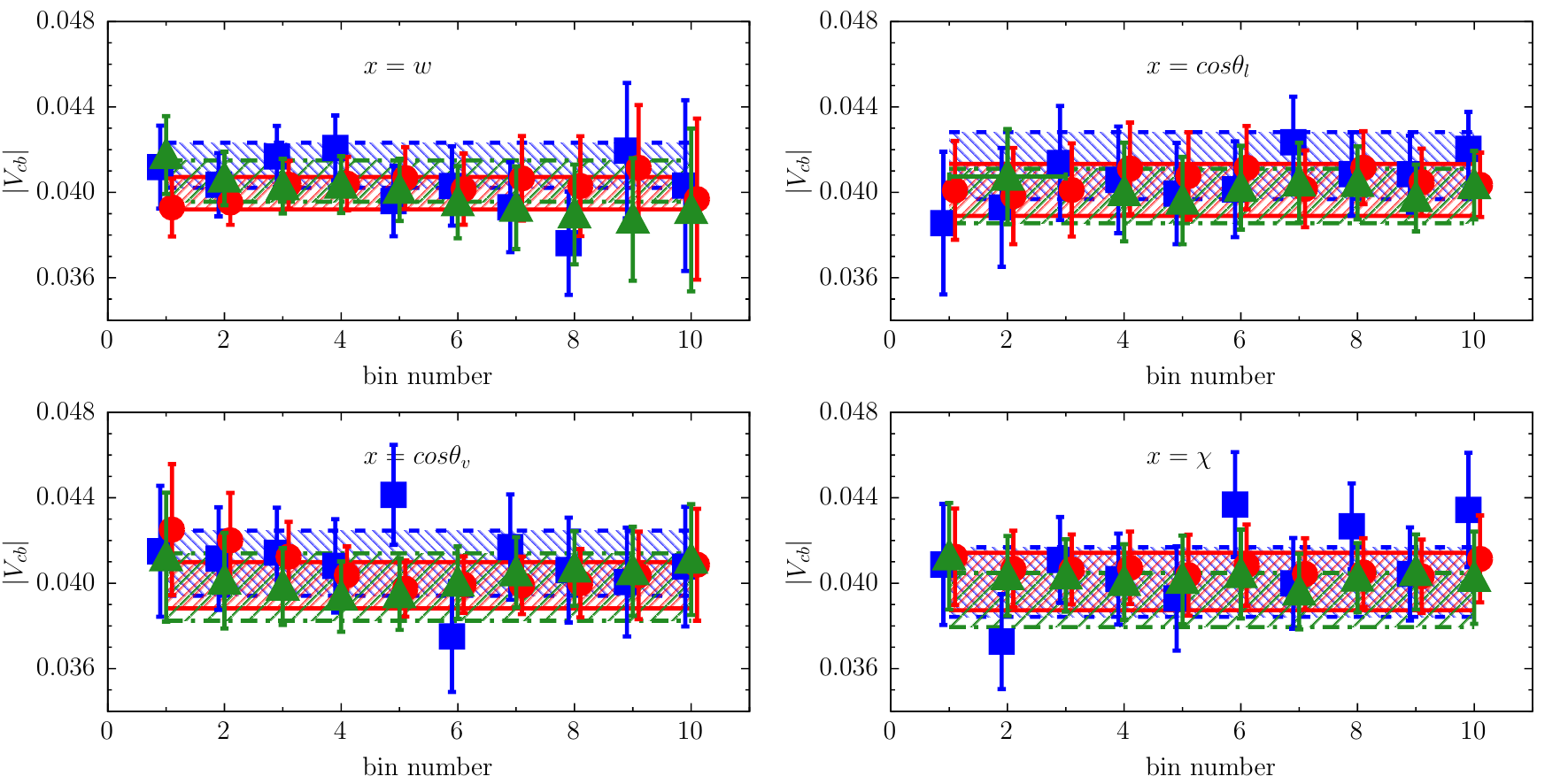}
\caption{\it \small The same as in Fig.\,\ref{fig:Vcb_origFNAL}, but using as inputs for the DM method  the lattice FFs evaluated by the JLQCD Collaboration\,\cite{Aoki:2023qpa}.}
\label{fig:Vcb_origJLQCD}
\end{center}
\end{figure}

For the sake of comparison, in Figs.~\ref{fig:Vcb_origFNAL}-\ref{fig:Vcb_origJLQCD} we show the bin-per-bin distributions of $\vert V_{cb} \vert$ for each kinematical variable $x$ and for each experiment.
Three important observations are necessary:
\begin{itemize}
\item we confirm the difficulty of the FFs by FNAL/MILC and HPQCD to  produce  the same value of $\vert V_{cb} \vert$ from different bins in $x$. This is a signal that  either the shape of the theoretical form factors suffer still uncorrected  systematic effects and/or that some other physical contribution is at work (systematic effects in  the data, contributions from physics beyond the Standard Model\,\cite{Fedele:2023ewe}, etc.);
\item in the top left panels the value of $ \vert V_{cb} \vert$ of Figs.\,\ref{fig:Vcb_origFNAL} and \ref{fig:Vcb_origHPQCD} exhibits a marked dependence on the specific $w$-bin. However, the value of $ \vert V_{cb} \vert$ obtained adopting a constant fit over the bins is dominated by the more precise bins at small values of the recoil, where either direct lattice data are available (FNAL/MILC) and the length of the momentum extrapolation is limited or systematic effects are expected to be smaller (HPQCD);
\item our understanding is that the problem above can be mostly attributed to the shape of $F_1(w)$ as a function of $w$, whereas the other form factors play a  minor role in this respect;
\item we remark, instead, the striking flatness of the values of $\vert V_{cb} \vert$ from JLQCD as a function of the bins in all the $x$ variables, mainly due to the different  shape of $F_1(w)$ as a function of $w$ with respect to the other collaborations. This is reflected also in better values of the reduced $\chi^2/(\mbox{d.o.f.})$ variable in the average of $\vert V_{cb} \vert$ over the different bins for each variable $x$ (see later Table\,\ref{tab:Vcb_orig});
\item as already noted in Ref.\,\cite{Martinelli:2021onb}, we confirm that the use of the original correlation matrix of Ref.\,\cite{Belle:2018ezy} would give rise to an anomalous underestimate of the mean values of $\vert V_{cb} \vert$ in the case of some of the variables $x$. This corresponds, however, also to much larger values of the reduced $\chi^2$-variable.  
\end{itemize}

Adopting the best constant fit over the $N _x$ bins we compute $|V_{cb}|_x$ and its variance $\,\sigma^2_{\vert V_{cb} \vert_x} $ for each kinematical variable $x$ and for each of the three experiments\,\cite{Belle:2023bwv, Belle:2018ezy, Belle-II:2023okj} as 
\bea
    \label{eq:muVcb}
    \vert V_{cb} \vert_x & = & \frac{\sum_{i, j=1}^{N_x} (\mathbf{C}_x^{-1})_{ij} \vert V_{cb} \vert_j}{\sum_{i, j=1}^{N_x} (\mathbf{C}_x^{-1})_{ij}} ~ , ~ \\[2mm]
    \label{eq:sigmaVcb}
    \,\sigma^2_{\vert V_{cb} \vert_x} & = & \frac{1}{\sum_{i, j=1}^{N_x} (\mathbf{C}_x^{-1})_{ij}} ~ , ~
\eea
where $\mathbf{C}_x$ is the $N_x \times N_x$ block of the full covariance matrix $\mathbf{C}$ (of dimension $N \times N$) corresponding to the kinematical variable $x$.
The results obtained for $|V_{cb}|_x$ and $\,\sigma_{\vert V_{cb} \vert_x}$ are collected in Table\,\ref{tab:Vcb_orig} together with the values of the reduced (correlated) $\chi^2/(\mbox{d.o.f.})$ variable, with the number of d.o.f.~being equal to $N_x - 1$.
\begin{table}[htb!]
\renewcommand{\arraystretch}{1.25}
\begin{center}
\begin{tabular}{|c||c|c|c|c|}
\hline
\multicolumn{5}{|l|}{\,\,\,\,\,\,\,\,\,\,\,\,\,\,\,\,\,\,\,\,\,\,\,\,\,\,\,\,\,\,\,\,\,\,\,\,\,\,\,\,\,\,\,\,\,\,\,\,\,\,\,\,\,\,\,\,\,\,\,\,\,\,\,\,\,\,\,\,\,\,\,\,\,\,\,\,\,\,\,\,\,\,\,\,\,\,\,\,\,\,\,\,\,\,\,\,\,\,\,\,\,\,\,\,\,\,\,\,\,input  FNAL/MILC}\\\hline
 experiment & ~$\vert V_{cb} \vert_{x = w}\cdot 10^{3}$ & ~$\vert V_{cb} \vert_{x = \mbox{cos}\theta_l}\cdot 10^{3}$~ & ~$\vert V_{cb} \vert_{x = \mbox{cos}\theta_v}\cdot 10^{3}$~ & ~$\vert V_{cb} \vert_{x = \chi}\cdot 10^{3}$~\\
\hline 
Belle '18\,\cite{Belle:2018ezy}           & ~39.4~(7)~ & ~40.9~(12)~ & ~40.0~(10)~ & ~42.7~(14)~\\
$\chi^2/(\mbox{d.o.f.})$                     & ~1.21~        &  ~1.36~         &~1.99~          &~0.38~ \\
\hline 
Belle '23\,\cite{Belle:2023bwv}          & ~40.2~(10)~ & ~42.9~(16)~ & ~42.6~(16)~ & ~42.5~(16)~\\
$\chi^2/(\mbox{d.o.f.})$                     & ~1.72~         &  ~0.83~         &~1.14~           &~1.94~ \\
\hline 
BelleII '23\,\cite{Belle-II:2023okj}      & ~39.3~(9)~ & ~40.5~(12)~ & ~41.1~(16)~ & ~41.1~(13)~\\
$\chi^2/(\mbox{d.o.f.})$                     & ~0.81~       &  ~2.55~         &~2.46~          &~1.36~ \\
\hline\hline
\multicolumn{5}{|l|}{\,\,\,\,\,\,\,\,\,\,\,\,\,\,\,\,\,\,\,\,\,\,\,\,\,\,\,\,\,\,\,\,\,\,\,\,\,\,\,\,\,\,\,\,\,\,\,\,\,\,\,\,\,\,\,\,\,\,\,\,\,\,\,\,\,\,\,\,\,\,\,\,\,\,\,\,\,\,\,\,\,\,\,\,\,\,\,\,\,\,\,\,\,\,\,\,\,\,\,\,\,\,\,\,\,\,\,\,\,input  HPQCD}\\\hline
 experiment & ~$\vert V_{cb} \vert_{x = w}\cdot 10^{3}$ & ~$\vert V_{cb} \vert_{x = \mbox{cos}\theta_l}\cdot 10^{3}$~ & ~$\vert V_{cb} \vert_{x = \mbox{cos}\theta_v}\cdot 10^{3}$~ & ~$\vert V_{cb} \vert_{x = \chi}\cdot 10^{3}$~\\
\hline 
Belle '18\,\cite{Belle:2018ezy}         & ~39.4~(9)~ & ~40.3~(15)~ & ~39.7~(13)~ & ~41.5~(17)~\\
$\chi^2/(\mbox{d.o.f.})$                   & ~0.53~        &  ~0.59~        &  ~0.96~       & ~0.32~ \\
\hline 
Belle '23\,\cite{Belle:2023bwv}        & ~41.0~(12)~ & ~41.5~(18)~ & ~41.1~(17)~ & ~41.0~(19)~\\
$\chi^2/(\mbox{d.o.f.})$                   & ~1.21~          & ~0.63~         & ~0.86~          & ~1.65~ \\
\hline 
BelleII '23\,\cite{Belle-II:2023okj}    & ~39.8 (11)~ & ~39.9~(15)~ & ~39.4~(18)~ & ~40.2~(16)~\\
$\chi^2/(\mbox{d.o.f.})$                   & ~0.37~        & ~1.63~         & ~1.52~         & ~1.15~ \\
\hline\hline
\multicolumn{5}{|l|}{\,\,\,\,\,\,\,\,\,\,\,\,\,\,\,\,\,\,\,\,\,\,\,\,\,\,\,\,\,\,\,\,\,\,\,\,\,\,\,\,\,\,\,\,\,\,\,\,\,\,\,\,\,\,\,\,\,\,\,\,\,\,\,\,\,\,\,\,\,\,\,\,\,\,\,\,\,\,\,\,\,\,\,\,\,\,\,\,\,\,\,\,\,\,\,\,\,\,\,\,\,\,\,\,\,\,\,\,\,input JLQCD}\\\hline
 experiment & ~$\vert V_{cb} \vert_{x = w}\cdot 10^{3}$ & ~$\vert V_{cb} \vert_{x = \mbox{cos}\theta_l}\cdot 10^{3}$~ & ~$\vert V_{cb} \vert_{x = \mbox{cos}\theta_v}\cdot 10^{3}$~ & ~$\vert V_{cb} \vert_{x = \chi}\cdot 10^{3}$~\\
\hline 
Belle '18\,\cite{Belle:2018ezy}       & ~40.0~(8)~ & ~40.1~(12)~ & ~39.9~(11)~ & ~40.1~(13)~\\
$\chi^2/(\mbox{d.o.f.})$                 & ~0.24~        &  ~0.24~        &~0.38~          &~0.10~ \\
\hline 
Belle '23\,\cite{Belle:2023bwv}      & ~41.3~(11)~ & ~41.3~(16)~ & ~40.9~(15)~ & ~40.0~(16)~\\
$\chi^2/(\mbox{d.o.f.})$                 & ~1.72~         &  ~0.50~         &~0.60~           &~1.69~ \\
\hline 
BelleII '23\,\cite{Belle-II:2023okj}  & ~40.5~(10)~ & ~39.8~(13)~ & ~39.8~(16)~ & ~39.2~(13)~\\
$\chi^2/(\mbox{d.o.f.})$                & ~0.62~          &  ~1.47~        &~1.41~           &~0.99~ \\
\hline
\end{tabular}
\end{center}
\renewcommand{\arraystretch}{1.0}
\caption{\it \small Mean values and uncertainties of the CKM element $\vert V_{cb} \vert$ obtained by the correlated average procedure given by Eqs.\,(\ref{eq:muVcb})-(\ref{eq:sigmaVcb}) for each of the four kinematical variables $x$ and for each of the three experimental data sets\,\cite{Belle:2023bwv, Belle:2018ezy, Belle-II:2023okj} and of the lattice inputs. The corresponding values of the reduced (correlated) $\chi^2/(\mbox{d.o.f.})$ variable are also shown.}
\label{tab:Vcb_orig}
\end{table}

Then, we evaluate the correlation matrix $\rho_{x x^\prime}$ among the values of $\vert V_{cb} \vert_x$ corresponding to the four kinematical variables $x$ for each of the three lattice inputs and for each experiment. A simple calculation yields
\be
    \rho_{x x^\prime} = \sigma_{\vert V_{cb} \vert_x} \sigma_{\vert V_{cb} \vert_{x^\prime}} \sum_{i_x, j_x = 1}^{N_x} ~ \sum_{i_{x^\prime}, j_{x^\prime} = 1}^{N_{x^\prime}} 
        (\mathbf{C}_x^{-1})_{i_x j_x }  (\mathbf{C}_{x^\prime}^{-1})_{i_{x^\prime} j_{x^\prime}} \mathbf{C}_{\tilde{j}_x \tilde{j}_{x^\prime}} ~ , ~
\ee
where $\tilde{j}_x$ corresponds to the index $j_x$ in the full $N \times N$ covariance matrix $\mathbf{C}$. 
We inflate the uncertainties $\,\sigma_{\vert V_{cb} \vert_x}$ by multiplying them by a PDG scale factor, namely we consider the following covariance matrix
\be
     \rho_{x x^\prime} \sigma_{\vert V_{cb} \vert_x} \sigma_{\vert V_{cb} \vert_{x^\prime}}  f_x f_{x^\prime} ~ , ~
\ee 
where $f \equiv \sqrt{\chi^2/(\mbox{d.o.f.})}$ when $\chi^2/(\mbox{d.o.f.}) > 1$ and $1$ otherwise\,\cite{ParticleDataGroup:2022pth}.
Finally, we apply the correlated procedure given by Eqs.\,(\ref{eq:muVcb})-(\ref{eq:sigmaVcb}) (with $N_x \to 4$), obtaining the results shown in columns 2-4 of Table\,\ref{tab:Vcb3}.
\begin{table}[htb!]
\renewcommand{\arraystretch}{1.25}
\begin{center}
\begin{tabular}{||c||c|c|c||c||}
\hline
\multicolumn{5}{|l|}{\,\,\,\,\,\,\,\,\,\,\,\,\,\,\,\,\,\,\,\,\,\,\,\,\,\,\,\,\,\,\,\,\,\,\,\,\,\,\,\,\,\,\,\,\,\,\,\,\,\,\,\,\,\,\,\,\,\,\,\,\,\,\,\,\,\,\,\,\,\,\,\,\,\,\,\,\,\,\,\,\,\,\,\,\,\,\,\,\,\,\,\,\,\,\,\, $\vert V_{cb} \vert \, \cdot 10^{3}$}\\\hline
 experiment & ~FNAL/MILC & ~HPQCD~ & ~JLQCD~& ~Average~\\
\hline 
Belle '18\,\cite{Belle:2018ezy}       & ~39.64~(74)~   & ~39.11~(81)~    & ~39.92~(74)~   & ~39.58~(98)~\\
$\chi^2/(\mbox{d.o.f.})$                 & ~3.71~              &  ~1.14~              &~0.04~              &~0.26~\\
\hline 
Belle '23\,\cite{Belle:2023bwv}      & ~40.87~(115)~ & ~41.03~(125)~ & ~41.38~(134)~ & ~41.11~(138)~\\
$\chi^2/(\mbox{d.o.f.})$                 & ~1.80~             &  ~0.11~             &~0.31~               &~0.03~\\
\hline 
BelleII '23\,\cite{Belle-II:2023okj}  & ~39.35~(77)~   & ~39.98~(102)~   & ~40.20~(85)~   & ~39.79~(94)~\\
$\chi^2/(\mbox{d.o.f.})$                 & ~0.63~             &  ~0.09~              &~0.42~               &~0.29~\\
 \hline
\end{tabular}
\end{center}
\renewcommand{\arraystretch}{1.0}
\caption{\it \small Mean values and uncertainties of the CKM element $\vert V_{cb} \vert$ obtained by  combining the correlated averages of the four kinematical variables $x$  for each of the three data sets \,\cite{Belle:2023bwv, Belle:2018ezy, Belle-II:2023okj} and for each of the lattice inputs. In the last column for each experimental data set we give the averages obtained using Eqs.\,(\ref{eq:average28})-(\ref{eq:sigma28}) with $w_k \propto \sigma_k^{-2}$ and including in the individual uncertainties the corresponding PDG scale factor.}
\label{tab:Vcb3}
\end{table}
In the last column, for each experiment we give the corresponding averages over the lattice inputs, obtained using Eqs.\,(\ref{eq:average28})-(\ref{eq:sigma28}) with $w_k \propto \sigma_k^{-2}$ and including in the individual uncertainties the corresponding PDG scale factor.

Finally, we combine the results of the last column of Table\,\ref{tab:Vcb3}, taking into account a correlation coefficient equal to $0.2$ between the two Belle determinations, obtaining for $|V_{cb}|$ the estimate
\be 
    \vert V_{cb} \vert = (39.92 \pm 0.64)\cdot 10^{-3} ~ , ~
    \label{eq:finalvcb} 
\ee
where the uncertainty includes a PDG scaling factor equal to $1.0$. 
Had we used Eqs.\,(\ref{eq:average28})-(\ref{eq:sigma28})  to combine the individual results of Table\,\ref{tab:Vcb_orig} we would have gotten $\vert V_{cb} \vert = (40.58 \pm 1.67 )\cdot 10^{-3}$ using $w_k = 1 / N$ and $\vert V_{cb} \vert = (40.30 \pm 1.50 )\cdot 10^{-3}$ using $w_k \propto \sigma_k^{-2}$.  
These findings confirm the trustability of Eq.\,(\ref{eq:finalvcb}).

The result\,(\ref{eq:finalvcb}) is in nice agreement with the one of Ref.\, \cite{Ray:2023xjn}, namely $\vert V_{cb} \vert = (40.3 \pm 0.5)\cdot 10^{-3}$, which was  obtained without including the Belle-II data set of Ref.\,\cite{Belle-II:2023okj} and the HPQCD FFs in the analysis.

\subsubsection{$\mathbf{\vert V_{cb} \vert}$ from a single analysis}
\label{sec:unos}

In this Section we apply the DM method to all the values of the FFs of the three lattice Collaborations\,\cite{FermilabLattice:2021cdg, Harrison:2023dzh, Aoki:2023qpa} as they were the results of a single lattice calculation. As shown in Figs.\,\ref{fig:FFsDM} and \ref{fig:FFsBGL}, for each of the four FFs we have 3 data points from Ref.\,\cite{FermilabLattice:2021cdg}, 3 data points from Ref.\cite{Aoki:2023qpa} and 4 data points from Ref.\,\cite{Harrison:2023dzh} for a total of 10 data points for each FF. Correlations among the FFs within the same lattice calculation are taken into account, whereas we consider independent the results of the three collaborations, since they are based on different sets of gauge configurations (this holds also for the lattice setups of Refs.\,\cite{FermilabLattice:2021cdg} and \cite{Harrison:2023dzh}).  

According to the DM method, using multivariate Gaussian distributions we generate a sample of events (of the order of $10^5$), each of which is composed by 40 data points for the FFs (10 points for each FF) plus 3 data points for the relevant nonperturbative susceptibilities, whose values are taken from Ref.\,\cite{Martinelli:2021frl}.

Typically, the DM unitary filters are satisfied only by a reduced number of the input events. The percentage of the surviving events turns out to be only $\approx 1\%$ already in the case of the FFs corresponding either to Ref.\,\cite{FermilabLattice:2021cdg} or Ref.\,\cite{Aoki:2023qpa}  (i.e., for a total of 12 data points in each case) and even less for the 16 data points of Ref.\,\cite{Harrison:2023dzh}. When all the FFs of the three lattice Collaborations are considered simultaneously, the DM unitary filters become extremely selective and no event satisfies the filters even when we use an initial sample of $10^6$ events or more. 

Therefore, in Ref.\,\cite{Simula:2023ujs} two of us have developed a unitary sampling procedure, which is basically a kind of importance sampling ($IS$), which easily allows to generate events for the FFs satisfying the unitary filters for any number of initial data points. The $IS$ procedure is described in Appendix\,\ref{sec:appA} for the present case of the $B \to D^*$ FFs, where we illustrate also how the KCs\,(\ref{eq:KCs}) are properly implemented. In what follows we denote the results obtained using the $IS$ procedure as the DM$_{IS}$ ones.

\begin{figure}[htb!]
\begin{center}
\includegraphics[scale=0.3]{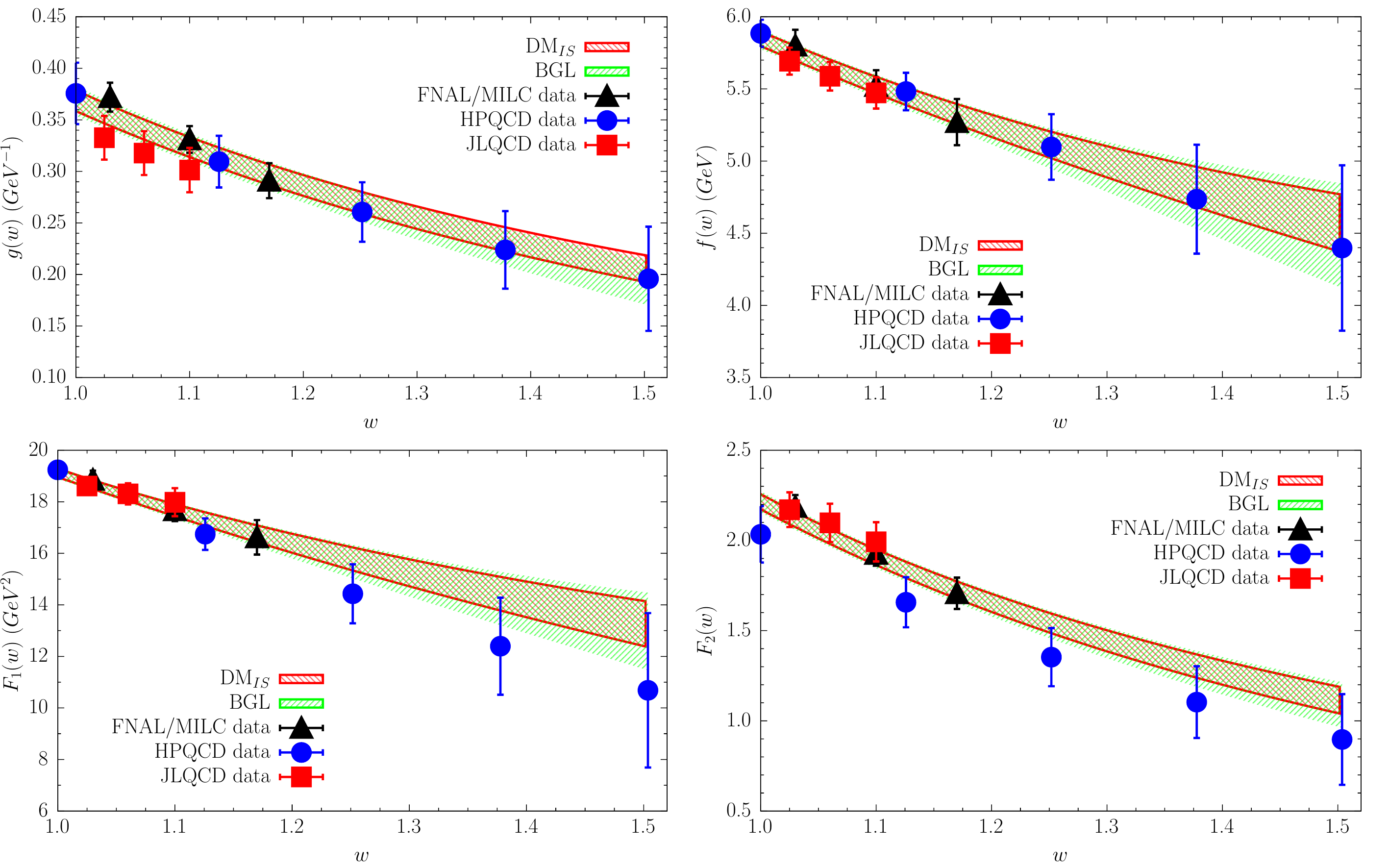}
\caption{\it \small The FFs from Refs.\,\cite{FermilabLattice:2021cdg, Harrison:2023dzh, Aoki:2023qpa} together with the bands (at $1\,\sigma$ level) obtained by using simultaneously all the lattice inputs within the $DM_{IS}$ method (red bands) or the BGL approach supplemented by the unitary and kinematical constraints (green bands). The BGL $z$-expansions are truncated after the quartic (quintic) term for the FFs $g$ and $f$ ($F_1$ and $F_2$). The DM bands are rigorously truncation independent.}
\label{fig:insieme1}
\end{center}
\end{figure}

In Fig.\,\ref{fig:insieme1} we show the bands obtained by using simultaneously the results of the three Collaborations\,\cite{FermilabLattice:2021cdg, Harrison:2023dzh, Aoki:2023qpa} within either the DM$_{IS}$ method (red bands) or the BGL approach (green bands). In the latter case the unitary and kinematical constraints are directly implemented in the minimization procedure of the correlated $\chi^2$-variable (see Ref.\,\cite{Simula:2023ujs}). The differences at high recoil are mainly produced in the BGL approach by the subset of input data which do not satisfy unitarity, whereas the DM method is free by construction from this problem.

\begin{figure}[htb!]
\begin{center}
\includegraphics[scale=0.50]{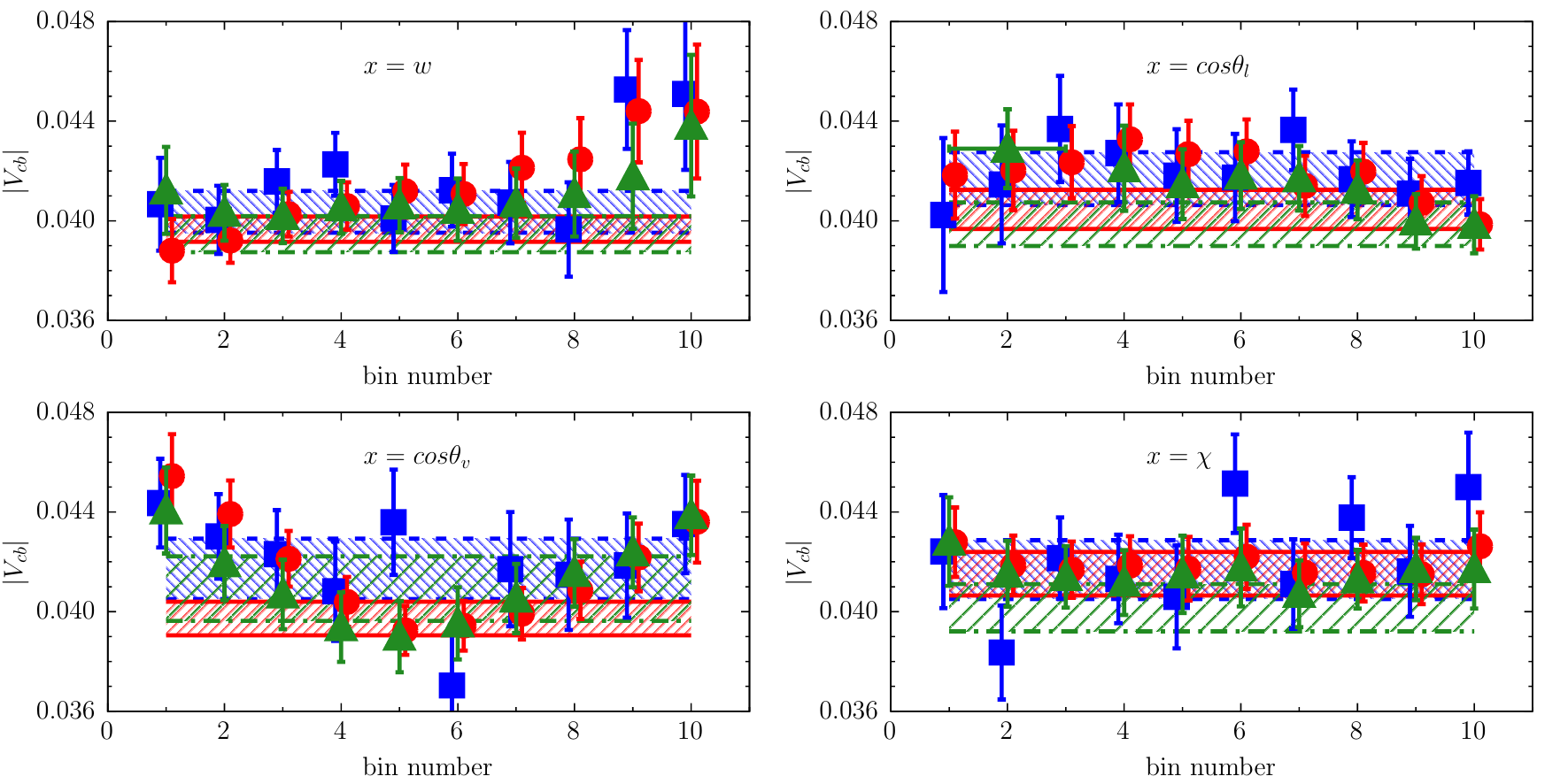}
\caption{\it \small The bin-per-bin estimates of $\vert V_{cb} \vert$ for each kinematical variable $x$ using the experimental data from Refs.\,\cite{Belle:2023bwv} (blue squares),\,\cite{Belle-II:2023okj} (green triangles) and \cite{Belle:2018ezy} (red circles), adopting the FFs extrapolated by the DM$_{IS}$ method applied to all the lattice results from Refs.\,\cite{FermilabLattice:2021cdg}-\cite{Aoki:2023qpa}.  The horizontal blue, green and red bands correspond to the correlated averages evaluated through Eqs.\,(\ref{eq:muVcb})-(\ref{eq:sigmaVcb}) over the bins and shown in Table~\ref{tab:Vcbtogether}.}
\label{fig:insieme}
\end{center}
\end{figure}

The binned values of $\vert V_{cb} \vert$ obtained from the differential distributions $d\Gamma / dx$ are shown in Fig.\,\ref{fig:insieme}.
Following the same procedure of the previous subsection, we evaluate the correlated averages through Eqs.\,(\ref{eq:muVcb})-(\ref{eq:sigmaVcb}) over the bins for each kinematical variable $x$. The results are collected in Table\,\ref{tab:Vcbtogether} together with their correlated averages for each experiment, evaluated including in the individual uncertainties the corresponding PDG scale factor.

\begin{table}[htb!]
\renewcommand{\arraystretch}{1.25}
\begin{center}
\begin{adjustbox}{max width=\textwidth}
\begin{tabular}{||c||c|c|c|c||c||}
\hline
\multicolumn{6}{|l|}{\,\,\,\,\,\,\,\,\,\,\,\,\,\,\,\,\,\,\,\,\,\,\,\,\,\,\,\,\,\,\,\,\,\,\,\,\,\,\,\,\,\,\,\,\,\,\,\,\,\,\,\,\,\,\,\,\,\,\,\,\,\,\,\,\,\,\,\,\,\,\,\,\,\,\,\,\,\,\,\,\,\,\,\,\,\,\,\,\,\,\,\,\,\,\,\, input FNAL/MILC + HPQCD + JLQCD}\\\hline
experiment & ~$\vert V_{cb} \vert_{x = w}\cdot 10^{3}$ & ~$\vert V_{cb} \vert_{x = \mbox{cos}\theta_l}\cdot 10^{3}$~ & ~$\vert V_{cb} \vert_{x = \mbox{cos}\theta_v}\cdot 10^{3}$~ & ~$\vert V_{cb} \vert_{x = \chi}\cdot 10^{3}$~& Average\\
\hline 
Belle '18\,\cite{Belle:2018ezy}       & ~39.66~(51)~ & ~40.46~(78)~   & ~39.72~(68)~  & ~41.52~(88)   & ~39.81~(50)~\\
$\chi^2/(\mbox{d.o.f.})$                 & ~0.76~            &  ~1.10~            & ~1.99~             & ~0.33~           & ~3.48~\\
\hline 
Belle '23\,\cite{Belle:2023bwv}      & ~40.36~(84)~ & ~41.69~(106)~ & ~41.72~(121)~ & ~41.69~(118) & ~41.13~(91)~\\
$\chi^2/(\mbox{d.o.f.})$                 & ~1.90~           &  ~0.62~             & ~1.07~             & ~1.84~           & ~1.00~\\
 \hline
BelleII '23\,\cite{Belle-II:2023okj}  & ~39.47~(72)~ & ~39.87~(87)~   & ~40.93~(129)~ & ~40.16~(95)   & ~39.38~(64)~\\
$\chi^2/(\mbox{d.o.f.})$                 & ~0.75~           &  ~2.52~            & ~2.54~             & ~1.39~            & ~0.29~\\
 \hline
\end{tabular}
\end{adjustbox}
\end{center}
\renewcommand{\arraystretch}{1.0}
\caption{\it \small Mean values and uncertainties of the CKM element $\vert V_{cb} \vert$ obtained by the correlated average procedure for each of the four kinematical variables $x$ and for each experiment\,\cite{Belle:2023bwv, Belle:2018ezy, Belle-II:2023okj}, using simultaneously the three lattice calculations of the FFs extrapolated by the DM$_{IS}$ approach (see text and Fig.\,\ref{fig:insieme}). In the last column for each experiment we give the correlated average of the four determinations, including in the individual uncertainties the corresponding PDG scale factor.}
\label{tab:Vcbtogether}
\end{table}

By combining the results of the last column of Table\,\ref{tab:Vcbtogether}, taking into account  in the individual uncertainties the corresponding PDG scale factor as well as a correlation coefficient equal to $0.2$ between the two Belle determinations, we obtain our estimate 
\be  
    \vert V_{cb} \vert \cdot 10^{3} = 39.87 \pm 0.55 ~ , ~
    \label{eq:together}
\ee
where the uncertainty includes a PDG scaling factor equal to $1.15$. The result\,(\ref{eq:together}) agrees very well with Eq.\,(\ref{eq:finalvcb}), obtained by averaging the results corresponding to the three independent analyses carried out in Section\,\ref{sec:trino}.

\subsection{Evaluation of $R(D^*)$ and polarization observables}
\label{sec:RDstar}

By using the DM bands of the FFs presented in Sections\,\ref{sec:trino} and \ref{sec:unos} we can compute the pure theoretical expectation values of the ratio $R(D^*)$, the $\tau$-polarization $P_{\tau}(D^*)$, the longitudinal $D^*$-polarization fraction for heavy and light charged leptons, $F_{L, \tau}$ and $F_{L, \ell}$ ($\ell = e, \mu$) respectively, and the forward-backward asymmetry $A_{FB, \ell}$. 
It is useful we collect the explicit formulae used for $F_{L, \tau}$, $F_{L, \ell}$ and $A_{FB, \ell}$ in terms of the helicity amplitudes $H_{0,\pm}(w)$ and $H_{0t}(w)$:
 \bea  
     \label{eq:FLtau} 
     F_{L, \tau} & = &  \frac{\int d[w ] \left[ |H_0(w)|^2  \left(1 + \frac{m_\tau^2}{2q^2} \right) + \frac{3}{2} |H_{0t}(w)|^2 \,\frac{m_\tau^2}{q^2} \right]}
                                 {\int d[w ] \left\{ \left[ |H_+(w)|^2 + |H_-(w)|^2 + |H_0(w)|^2 \right] \left(1 + \frac{m_\tau^2}{2 q^2} \right) + 
                                 \frac{3}{2} |H_{0t}(w)|^2 \,\frac{m_\tau^2}{q^2} \right\}} ~ , ~\\[2mm]
     \label{eq:FL}
     F_{L, \ell} & = & \frac{\int d[w] |H_0(w)|^2}{\int d[w] \left[ |H_0(w)|^2 + |H_+(w)|^2 + |H_-(w)|^2 \right]} ~ , ~ \\[2mm]
     \label{eq:AFBl} 
     A_{FB, \ell} & = & \frac{3}{4} \frac{\int d[w] \left[ |H_-(w)|^2 - |H_+(w)|^2\right]}{\int d[w] \left[ |H_0(w)|^2 + |H_+(w)|^2 + |H_-(w)|^2 \right]} ~ , ~ 
 \eea
where $d[w]$ is the phase space integration volume given by  
\be
    d[w] = q^2 \, m_{D^*} \sqrt{w^2 - 1} \left(1 - \frac{m_\tau^2}{q^2} \right)^2 dw ~ . ~ \nonumber 
\ee
and
\bea
     H_\pm(w) & = & f(w) \pm m_B^2 r \sqrt{w^2 - 1} ~ g(w) ~, ~ \nonumber \\[2mm]
     H_0(w) & = & \frac{F_1(w)}{\sqrt{q^2}} ~ , ~ \nonumber \\[2mm]
     H_{0t}(w) & = & m_B^2 r \sqrt{w^2 - 1} ~ \frac{F_2(w)}{\sqrt{q^2}} ~ . ~ \nonumber
\eea
 
The theoretical predictions are given in Table\,\ref{tab:rds} and they have been obtained either by computing the quantities above by using the FFs from the three collaborations separately, by averaging the three separate results according to the PDG procedure\,\cite{ParticleDataGroup:2022pth} (Average) and by combining the FFs of the three collaborations as they were the results of a single calculation using the $DM_{IS}$  approach (Combined). The two different procedures give well compatible results within the errors, which are larger for the Average due to the PDG scale factors.
\begin{table}[htb!]
\renewcommand{\arraystretch}{1.25}
\begin{center}
\begin{adjustbox}{max width=\textwidth}
\begin{tabular}{|c|c|c|c|c|c|}
\hline
Lattice FFs & ~$R(D^*)$~  & ~$P_{\tau}(D^*)$~ & ~$F_{L, \tau}$~ & ~$F_{L, \ell}$ & ~$A_{FB, \ell}$~ \\
\hline \hline
FNAL/MILC\,\cite{FermilabLattice:2021cdg} & 0.275(8) & -0.529(7) & 0.418(9)& 0.450(19)& 0.261(14) \\
\hline 
HPQCD\,\cite{Harrison:2023dzh} & 0.266(12) & -0.543(18) & 0.399(23) & 0.435(42)& 0.265(30)\\
\hline 
JLQCD\,\cite{Aoki:2023qpa} & 0.247(8) & -0.509(11) & 0.448(16) & 0.516(29)& 0.220(21) \\
\hline \hline
Average\,\cite{FermilabLattice:2021cdg}-\cite{ Aoki:2023qpa} & 0.262(9) & -0.525(7) & 0.422(10) & 0.465(22) & 0.251(13) \\
(PDG scale factor) & (1.8) & (1.3) & (1.4) & (1.5) & (1.2) \\
\hline 
Combined\,\cite{FermilabLattice:2021cdg}-\cite{ Aoki:2023qpa} & 0.259(5) & -0.521(6) & 0.425(7)& 0.473(14)& 0.252(10) \\
\hline \hline
Experimental value & 0.284(12)\,\cite{HeavyFlavorAveragingGroup:2022wzx} & -$0.38\pm 0.51^{+0.21}_{-0.16}$\,\cite{Belle:2016dyj} & 0.49(8)\,\cite{Belle:2019ewo, LHCb:2023ssl} & 0.520(6)\,\cite{Belle:2023bwv, Belle-II:2023okj} & 0.232(10)\,\cite{Belle:2023bwv, Belle-II:2023okj} \\
\hline 
\end{tabular}
\end{adjustbox}
\end{center}
\renewcommand{\arraystretch}{1.0}
\caption{\it \small Theoretical predictions for the $\tau / \mu$ ratio $R(D^*)$, the $\tau$-polarization $P_{\tau}(D^*)$, the longitudinal $D^*$-polarization fractions $F_{L,\tau}$ and $F_{L,\ell}$, and the forward-backward asymmetry $A_{FB,\ell}$ using the FFs from the three lattice Collaborations\,\cite{FermilabLattice:2021cdg, Harrison:2023dzh, Aoki:2023qpa}. We also give the average of the three separate results according to the PDG procedure\,\cite{ParticleDataGroup:2022pth} (Average), including the scale factor for the uncertainty, and the results obtained by combining the FFs of the three lattice Collaborations as they were the results of a single calculation using the $DM_{IS}$ approach (Combined). We also provide experimental values and uncertainties (see text).}
\label{tab:rds}
\end{table}

The experimental value of $F_{L, \tau}$ shown in the fourth column of Table\,\ref{tab:rds} has been obtained by averaging the results $F_{L, \tau} = 0.60(8) (4)$ from Ref.\,\cite{Belle:2019ewo} and $F_{L, \tau} = 0.43(6) (3)$ from Ref.\,\cite{LHCb:2023ssl}.
The experimental value of  $F_{L, \ell}$ shown in the fifth column of Table\,\ref{tab:rds} has been obtained by averaging the following experimental numbers:
$F_{L, e} = 0.485(17) (5)$ and $F_{L, \mu} = 0.518 (17)(5)$ from Ref.\,\cite{Belle:2023bwv}, and $F_{L, e} = 0.520(5) (5 )$ and $F_{L, \mu} = 0.527 (5 )(5)$ from Ref.\,\cite{Belle-II:2023okj}.
The experimental value of $A_{FB,\ell}$ shown in the sixth column of Table\,\ref{tab:rds} has been obtained from the average of $A_{FB, e} = 0.230(18) (5)$ and 
$A_{FB ,\mu} = 0.252(19)(5)$ from Ref.\,\cite{Belle:2023bwv}, and $A_{FB, e} = 0.228(12)(18)$ and  $A_{FB, \mu } = 0.211 (11)(21)$ from Ref.\,\cite{Belle-II:2023okj}.

In Ref.\,\cite{LHCb:2023ssl} the LHCb Collaboration has measured the  longitudinal $D^*$-polarization fraction $F_{L,\tau}$ not only in the whole kinematical range, but also in two different $q^2$-bins, namely: $q^2 < 7$ GeV$^2$ (low-$q^2$) and $q^2 > 7$ GeV$^2$ (high-$q^2$). Our theoretical predictions are shown in Table\,\ref{tab:FLtau} and compare positively with the experimental results.
\begin{table}[htb!]
\renewcommand{\arraystretch}{1.25}
\begin{center}
\begin{tabular}{|c|c|c|}
\hline
Lattice FFs & ~low-$q^2$ bin~ & ~high-$q^2$ bin ~ \\
\hline \hline
FNAL/MILC\,\cite{FermilabLattice:2021cdg} & 0.486(15)& 0.381(5)~ \\
\hline 
HPQCD\,\cite{Harrison:2023dzh} & 0.459(38)& 0.367(14)\\
\hline 
JLQCD\,\cite{Aoki:2023qpa} & 0.534(25)& 0.398(10) \\
\hline \hline
Average\,\cite{FermilabLattice:2021cdg}-\cite{ Aoki:2023qpa} & 0.495(17) & 0.383(6)~ \\
(PDG scale factor) & (1.4) & (1.4) \\
\hline 
Combined\,\cite{FermilabLattice:2021cdg}-\cite{ Aoki:2023qpa} & 0.498(12)& 0.384(4)~ \\
\hline \hline
Experimental value\,\cite{LHCb:2023ssl} & 0.51(7)(3) & 0.35(8)(2) \\
\hline 
\end{tabular}
\end{center}
\renewcommand{\arraystretch}{1.0}
\caption{\it \small Longitudinal $D^*$-polarization fraction $F_{L,\tau}$ measured by LHCb\,\cite{LHCb:2023ssl} in two different $q^2$-bins: $q^2 < 7$ GeV$^2$ (low-$q^2$) and $q^2 > 7$ GeV$^2$ (high-$q^2$). The theoretical predictions correspond to the use of the FFs from the three lattice Collaborations\,\cite{FermilabLattice:2021cdg, Harrison:2023dzh, Aoki:2023qpa}. We also give the average of the three separate results according to the PDG procedure\,\cite{ParticleDataGroup:2022pth} (Average), including the scale factor for the uncertainty, and the results obtained by combining the FFs of the three lattice Collaborations as they were the results of a single calculation using the $DM_{IS}$ approach (Combined).}
\label{tab:FLtau}
\end{table}

For completeness, we have evaluated an improved version of the $\tau / \mu$ ratio proposed in Ref.\,\cite{Isidori:2020eyd} to try to minimize the impact of the uncertainties of the theoretical FFs, namely
\be
    R^{\rm opt}(D^*) \equiv \frac{\int_{m_\tau^2}^{q_{max}^2} dq^2 \frac{d\Gamma}{dq^2}(B \to D^* \tau \nu_\tau)}
                                   {\int_{m_\tau^2}^{q_{max}^2} dq^2 \left[ \frac{\omega_\tau(q^2)}{\omega_{e(\mu)}(q^2)} \right] 
                                   \frac{d\Gamma}{dq^2}(B \to D^* e(\mu) \nu_{e(\mu)})} ~ , ~                              
\ee
where $\omega_\ell(q^2) = (1 - m_\ell^2 / q^2)^2 (1 + m_\ell^2 / 2q^2)$. Using the lattice data from the FNAL/MILC\,\cite{FermilabLattice:2021cdg}, HPQCD\,\cite{Harrison:2023dzh} and JLQCD\,\cite{Aoki:2023qpa} Collaborations we get $R^{\rm opt}(D^*) =  1.0794 \pm 0.0046$, $1.0696 \pm 0.0119$, $1.0916 \pm 0.0073$, respectively. These results yield a PDG-averaged value $R^{\rm opt}(D^*) = 1.0815 \pm 0.0046$ with a scale factor equal to $\simeq 1.3$. Adopting the DM$_{IS}$ bands for the FFs we obtain $R^{\rm opt}(D^*) = 1.0834 \pm 0.0038$.
  
We consider as our best (conservative) determinations the results corresponding to the procedure labelled Average in Tables\,\ref{tab:rds} and \ref{tab:FLtau}, namely
\bea
    \label{eq:Rds}
    R(D^*) & = & +0.262 \pm 0.009 ~ , \qquad 1.5\,\sigma ~ , ~ \\[2mm]
    \label{eq:Ptau}
    P_{\tau}(D^*) & = & -0.525 \pm 0.007 ~ , \qquad 0.3\,\sigma ~ , ~ \\[2mm]
    \label{eq:FLt}
     F_{L ,\tau} & = & +0.422 \pm 0.010 ~ , \qquad 0.8\,\sigma ~ , ~ \\[2mm]
    \label{eq:FLt_lowq2}
     F_{L ,\tau}(q^2 < 7 ~\mbox{GeV}^2) & = & +0.495 \pm 0.017 ~ , \qquad 0.2\,\sigma ~ , ~ \\[2mm]
    \label{eq:FLt_highq2}
     F_{L ,\tau}(q^2 >7 ~\mbox{GeV}^2) & = & +0.383 \pm 0.006 ~ , \qquad 0.4\,\sigma ~ , ~ \\[2mm]
     \label{eq:FLl}
    F_{L, \ell} & = & +0.465 \pm 0.022 ~ , \qquad 2.4\,\sigma ~ , ~ \\[2mm]
   \label{eq:AFB}
    A_{FB ,\ell}& = & +0.251 \pm 0.013  ~ , \qquad 1.2\,\sigma ~ , ~ \\[2mm]
   \label{eq:Rds_opt}
   R^{\rm opt}(D^*) & = & 1.0815 \pm 0.0046 ~ . ~
\eea
For all the above quantities, but $ F_{L, \ell}$, the theoretical and the experimental values agree quite well within the uncertainties.
Note that the $R(D^*)$ anomaly results to be smaller with respect to the $\simeq 2.2\,\sigma$ tension stated recently by HFLAV Collaboration~\cite{HFLAV_web23} as well as in Refs.\,\cite{Ray:2023xjn, Cui:2023jiw}.

In Appendix\,\ref{sec:appD} we have collected the full covariance matrices of the set of pure theoretical observables
\be
     \{ \Gamma_\tau / |V_{cb}|^2, \Gamma_\ell / |V_{cb}|^2, R(D^*), P_\tau(D^*), F_{L, \tau}, F_{L, \ell}, A_{FB, \ell} \} ~ , ~ 
     \label{eq:set_cov}
\ee
where $\Gamma_\tau \equiv \Gamma(B \to D^* \tau \nu_\tau)$ is the total decay rate into $\tau$ leptons and $\Gamma_\ell \equiv \Gamma(B \to D^* \ell \nu_\ell)$ the one into the light leptons $\ell = e, \mu$, corresponding to the lattice FFs obtained via the DM method applied separately to each of the three lattice calculations of Refs.\,\cite{FermilabLattice:2021cdg, Harrison:2023dzh, Aoki:2023qpa} and to all the lattice results from the three Collaborations. Our results can be useful for phenomenological studies.

\begin{figure}[htb!]
\begin{center}
\includegraphics[scale=0.625]{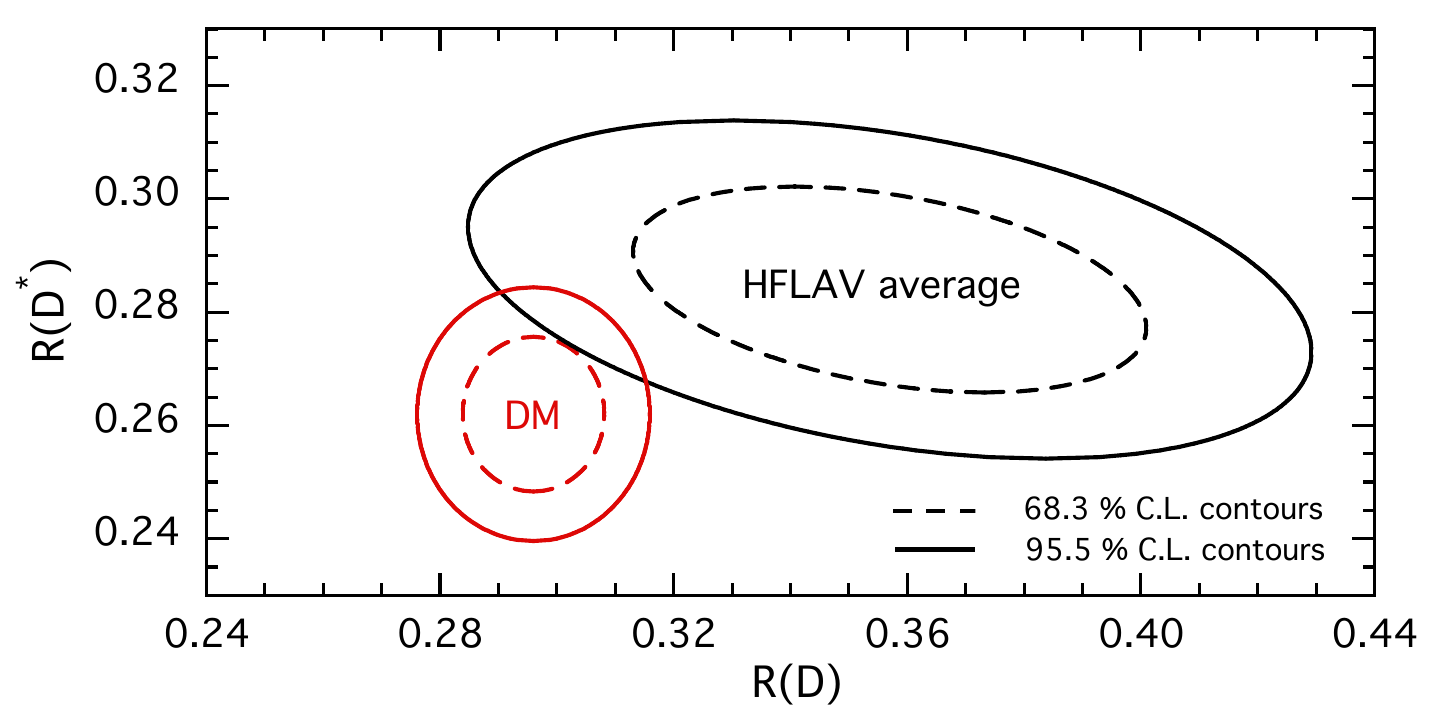}
\caption{\it \small The contour plots of the theoretical ratios $R(D)$ and $R(D^*)$, obtained by the DM method respectively in Ref.~\cite{Martinelli:2021onb} ($R(D) = 0.296 \pm 0.008$) and in this work (see Eq.\,(\ref{eq:Rds})), compared with those corresponding to the latest experimental world averages from HFLAV~\cite{HFLAV_web23} as for Summer 2023, namely $R(D)\vert_{\rm{exp}} = 0.357 \pm 0.029$ and $R(D^*)\vert_{\rm{exp}} = 0.284 \pm 0.012$.}
\label{fig:ellipse}
\end{center}
\end{figure}

In Ref.~\cite{Martinelli:2021onb} the DM method was applied to the final lattice data for the $B \to D \ell \nu_\ell$ transition provided by the FNAL/MILC Collaboration~\cite{MILC:2015uhg}. We obtained for the ratio $R(D)$ the pure theoretical estimate $R(D) = 0.296 \pm 0.008$, which is compatible with the latest experimental world average $R(D)\vert_{\rm{exp}} = 0.357 \pm 0.029$~\cite{HFLAV_web23} at the $\simeq 2.0\,\sigma$ level.
In Fig.~\ref{fig:ellipse} we show the comparison of the DM results for the two ratios $R(D)$ and $R(D^*)$ with the corresponding experimental world averages from HFLAV~\cite{HFLAV_web23}.
Note that we have considered our values for $R(D)$ and $R(D^*)$ as uncorrelated. This is motivated by the following considerations: ~ i) the lattice FFs entering the $B \to D$ decays share the same gauge configurations of only one of the three sets of lattice FFs adopted for the $B \to D^*$ decays; ~ ii) we expect that the $B \to D$ FFs may be correlated mainly with the $B \to D^*$ FF $g$, because all of them correspond to matrix elements of the vector weak current, while the other three $B \to D^*$ FFs are related  to matrix elements of the axial weak current; ~ iii)  we have explicitly checked that the correlation induced by the use of the vector transverse susceptibility $\chi_{1^-}(0)$, present in both decay channels, is very mild.

\section{A reappraisal of the calculation of $\vert V_{ub} \vert / \vert V_{cb} \vert$ from ${\cal{BR}}(B_s \to K \ell \nu_\ell) /$ ${\cal{BR}}(B_s \to D_s \ell \nu_\ell)$}
\label{sec:vubsvcb}

Stimulated by a new  lattice calculation of the FFs relevant for $B_s \to K \ell \nu$ decays\,\cite{Flynn:2023nhi}, we present in this Section a detailed discussion of the determination of the ratio $\vert V_{ub} \vert / \vert V_{cb} \vert$ and of its uncertainty.

The most precise value of $\vert V_{ub} \vert $ comes from $B \to \pi \ell  \nu_\ell$ semileptonic decays. By combining the values of $\vert V_{ub} \vert $ extracted from these processes and  from $B \to \tau \nu_\tau$ decays FLAG quotes\,\cite{FLAG_web23} 
\be  
    \vert V_{ub} \vert  =  (3.64 \pm 0.16 ) \cdot 10^{-3} ~ . ~ 
    \label{eq:vubflag} 
\ee

Nonetheless, a complementary and independent information on the exclusive value of $\vert V_{ub} \vert$, although with a much larger error, is  provided by the semileptonic $B_s \to K \ell \nu_\ell$ decay. 
This process, thanks to the experimental value of the ratio ${\cal{BR}}(B_s \to K \ell \nu_\ell) / {\cal{BR}}(B_s \to D_s \ell \nu_\ell)$, also provides a direct determination of $\vert V_{ub} \vert / \vert V_{cb} \vert$.

Both $\vert V_{ub} \vert $ and $\vert V_{ub} \vert / \vert V_{cb} \vert$ can indeed be determined from the ratio
\be 
    R_{BF} = \frac{{\cal{BR}}(B_s \to K \ell \nu_\ell)}{{\cal{BR}}(B_s \to D_s \ell \nu_\ell)} ~ 
\label{eq:RBF} 
\ee
using the experimental measurement of the branching ratio of the semileptonic decay $B_s \to D^-_s \mu^+ \nu_\mu$ and the $B_s$-meson lifetime, namely\,\cite{ParticleDataGroup:2022pth}
\be 
    {\cal{BR}}^{\rm exp}(B_s \to D^-_s \mu^+ \nu_\mu) = 2.44 (21) (10) \cdot 10^{-2}  ~ , \qquad \tau_{B_s} = 1.521 (5) \cdot 10^{-12} ~ s ~ .  ~
\ee
The experimental information on $R_{BF}$ is available for two separate bins in $q^2$ and in the whole kinematical range\,\cite{LHCb:2020ist}:
\bea  
    R^{(i=1=low)}_{BF} & = & 1.66(08)(09)  \cdot10^{-3} \qquad  q^2 \le 7 {\rm GeV}^2 \, , \nonumber \\[2mm]
    R^{(i=2=high)}_{BF} & = & 3.25(21)(19)  \cdot 10^{-3} \qquad  q^2 \ge 7 {\rm GeV}^2 \, , \\[2mm]
    R^{(i=3=total)}_{BF} & = & 4.89(21)(25)  \cdot 10^{-3} \, . \nonumber 
 \eea
In what follows we will make use of all the bins $i = 1, 2, 3$ separately, since they produce different results for $\vert V_{ub} \vert $ and $\vert V_{ub} \vert / \vert V_{cb} \vert$.
 
The value of $\vert V_{ub} \vert $ is extracted from the formula 
\be 
    \vert V_{ub} \vert^{(i)} =  \sqrt{R_{BF}^{(i)} \frac{{\cal{BR}}^{ exp}(B_s \to D^-_s \mu^+ \nu_\mu)}{\tau_{B_s } \widetilde{\Gamma}^{(i)}(B_s \to K \ell \nu_\ell)}} 
                                           \qquad { i}=1,2,3 ~ , ~
     \label{eq:vub}
\ee
where $\widetilde{\Gamma}^{(i)}(B_s \to K \ell \nu_\ell) = \Gamma^{(i)}(B_s \to K \ell \nu_\ell) / \vert V_{ub} \vert^2$ is the reduced theoretical rate, integrated over the $q^2$-range of the $i$-th bin and computed using lattice FFs.
 
The ratio $\vert V_{ub} \vert / \vert V_{cb} \vert$ is extracted from the expression 
\be 
     \frac{\vert V_{ub} \vert }{\vert V_{cb} \vert}^{(i)} = \sqrt{R_{BF}^{(i)} \frac{\widetilde{\Gamma}(B_s \to D_s \ell \nu_\ell)}{\widetilde{\Gamma}^{(i)}(B_s \to K \ell \nu_\ell)}} \, , 
     \label{eq:vubsvcb}
\ee
where $\widetilde{\Gamma}(B_s \to D_s \ell \nu_\ell) =  \Gamma(B_s \to D_s \ell \nu_\ell )/ \vert V_{cb} \vert^2$ is the reduced theoretical rate of the $B_s \to D_s \ell \nu_\ell$ process, computed using lattice FFs. 
We use $\widetilde{\Gamma}(B_s \to D_s \ell \nu_\ell) = (9.15 \pm 0.37 )\cdot 10^{12} s^{-1}$ taken from the latest FLAG review\,\cite{FlavourLatticeAveragingGroupFLAG:2021npn}.

\begin{figure}[htb!]
\centering
\includegraphics[width=\linewidth]{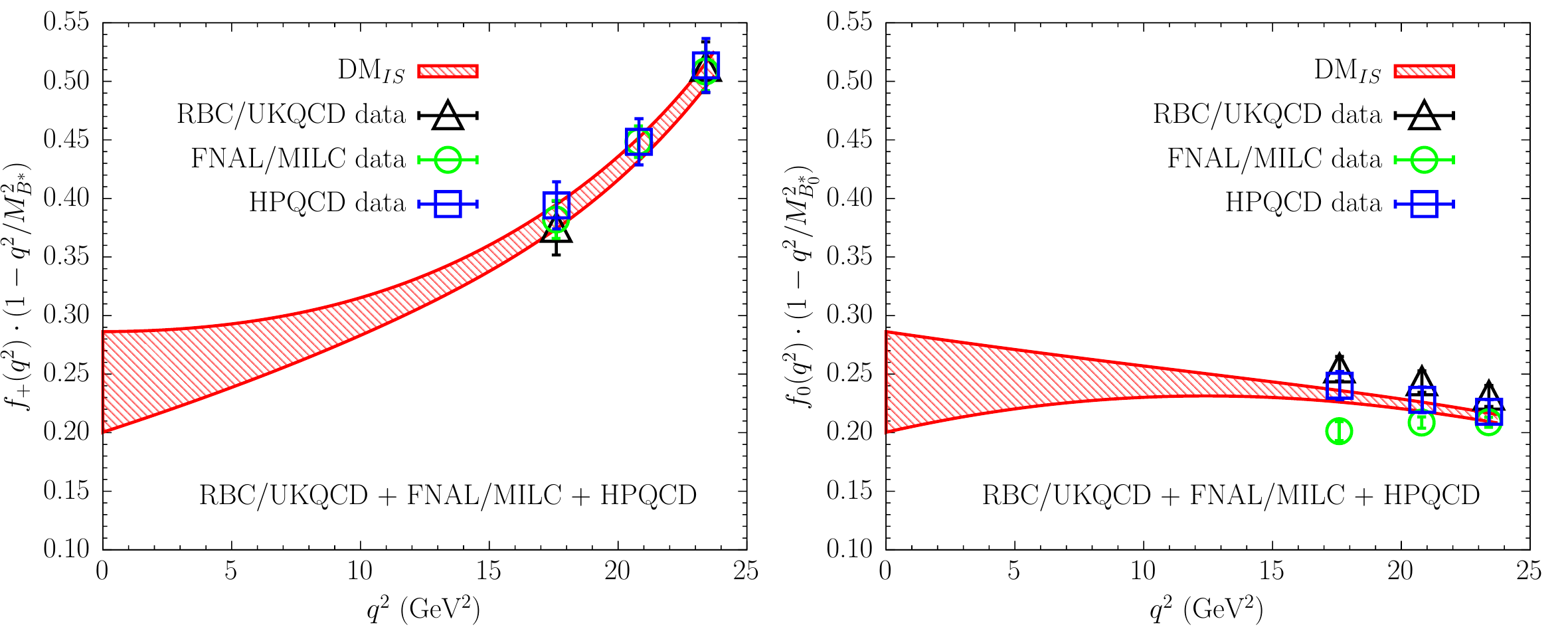} \\
\includegraphics[width=\linewidth]{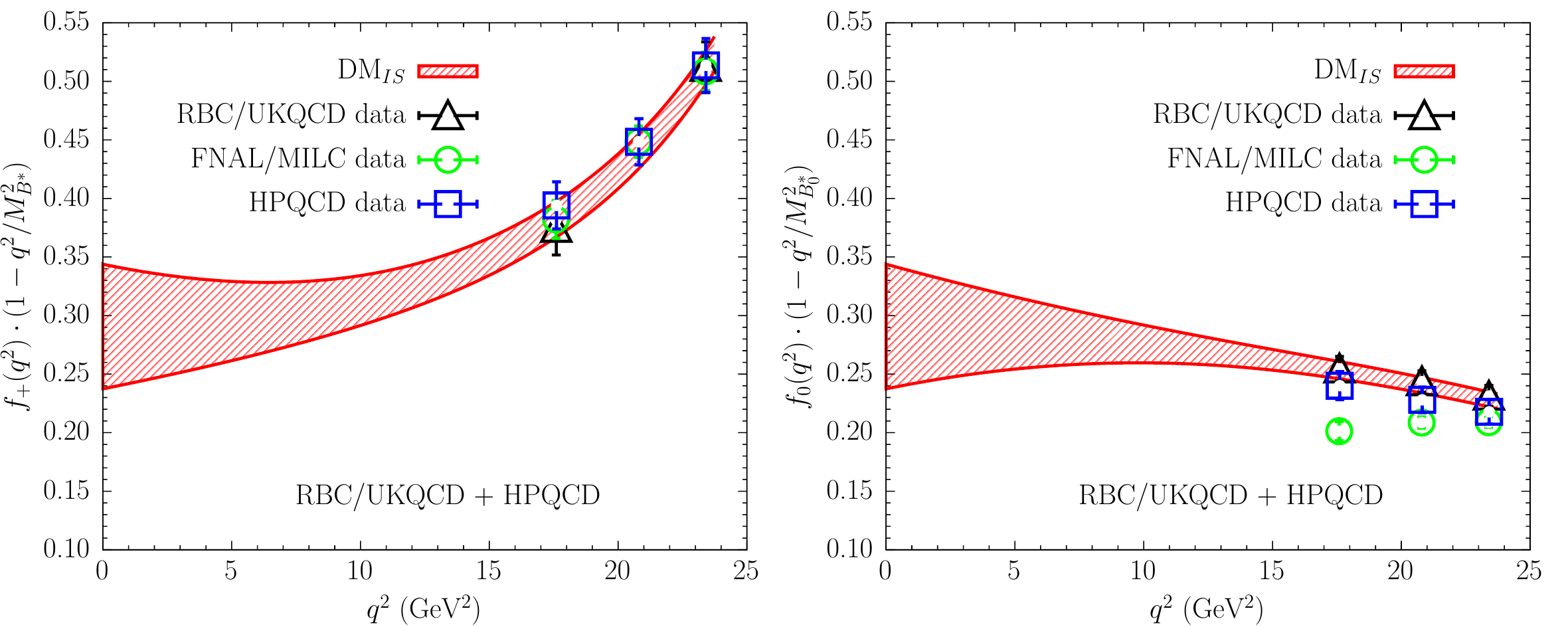}
\caption{\it \small Available lattice results for the vector $f_+(q^2)$ (left panels) and scalar $f_0(q^2)$ (right panels) FFs relevant for the $B_s \to K \ell \nu_\ell$ decays together with the $DM_{IS}$ bands obtained using the results of all the three collaborations RBC/UKQCD\,\cite{Flynn:2023nhi}, FNAL/MILC\,\cite{FermilabLattice:2019ikx} and HPQCD\,\cite{Bouchard:2014ypa, Monahan:2018lzv}  (top panels), or only the results from RBC/UKQCD\,\cite{Flynn:2023nhi} and HPQCD\,\cite{Bouchard:2014ypa, Monahan:2018lzv} (bottom panels). In the left panels the vector FF is multiplied by the factor $(1 - q^2 / m_{B^*}^2)$ with $m_{B^*} = 5.325$ GeV, while in the right panels the scalar FF is multiplied by the factor $(1 - q^2 / m_{B_0^*}^2)$ with $m_{B_0^*} = 5.68$ GeV (see Ref.\,\cite{Martinelli:2022tte}).}
\label{fig:FFBsK}
\end{figure}

In order to compute the r.h.s.~of Eqs.\,(\ref{eq:vub}) and (\ref{eq:vubsvcb}), besides the experimental inputs and the value of $\widetilde{\Gamma}(B_s \to D_s \ell \nu_\ell)$ from Ref.\,\cite{FlavourLatticeAveragingGroupFLAG:2021npn}, we need the vector and scalar FFs, $f_+(q^2)$ and $f_0(q^2)$, relevant for the theoretical prediction of $\widetilde{\Gamma}^{(i)}(B_s \to K \ell \nu_\ell)$.  
At present there are three independent lattice calculations of these FFs, the most recent one\,\cite{Flynn:2023nhi} and other two by FNAL/MILC\,\cite{FermilabLattice:2019ikx} and HPQCD\,\cite{Bouchard:2014ypa, Monahan:2018lzv}.
The lattice results, which are available only for $q^2 \gtrsim 17$ GeV$^2$, are shown in Fig.\,\ref{fig:FFBsK}. 
As already pointed out in Ref.\,\cite{Flynn:2023nhi}, there is a large difference in the case of the scalar FF $f_0(q^2)$ obtained by FNAL/MILC\,\cite{FermilabLattice:2019ikx} and by the other two Collaborations. This difference has consequences in the determination of $\vert V_{ub} \vert$ and $\vert V_{ub} \vert / \vert V_{cb} \vert$. 

For the definition of the conformal variable $z$ (see Eq.\,(\ref{eq:z}) in Appendix\,\ref{sec:DM}) and of the relevant kinematical functions we have followed Ref.\,\cite{Martinelli:2022tte}. In particular, for the nonperturbative dispersive bounds we have adopted the susceptibilities calculated in Ref.\,\cite{Martinelli:2022tte} at vanishing four-momentum transfer, namely $\chi_{0^+}(q_0^2 = 0) = (2.04 \pm 0.20) \cdot 10^{-2}$ and $\chi_{1^-}(q_0^2 = 0) = (4.45 \pm 1.16) \cdot 10^{-4}$ GeV$^{-2}$.
Notice that with our choice of the conformal variable $z$ branch points related to multiparticle production may occur inside the unit circle $|z| = 1$. However, following the approach of Ref.\,\cite{Boyd:1995sq}, we have verified that their impact on the dispersive bounds, as well as on the input data of the semileptonic FFs, is expected to be small and well within the uncertainties.
 
In Fig.\,\ref{fig:FFBsK} we show the extrapolated bands obtained by the DM$_{IS}$ approach using either all the lattice results from the three Collaborations as they were a single lattice calculation (top panels) or only the FFs from RBC/UKQCD and HPQCD (bottom panels). At $q^2 = 0$ the common value $f(0) = f_0(0) = f_+(0)$ turns out to be different, but consistent within the uncertainties, namely
\bea
    \label{eq:f0_DM}
    f^{DM_{IS}}(0) & = &  0.243 \pm 0.043 \qquad  \qquad {\rm FNAL/MILC+HPQCD +RBC/UKQCD} ~ , ~ \nonumber \\[2mm]
    f^{DM_{IS}}(0) & = &  0.291 \pm 0.053 \qquad  \qquad {\rm HPQCD +RBC/UKQCD} \nonumber ~ . ~ 
\eea
We have also repeated the extrapolations of the FFS adopting the BGL\,\cite{Boyd:1997kz} approach complemented by the unitary and kinematical constraints according to the procedure described in Ref.\,\cite{Simula:2023ujs}. We get results for $f(0)$ lower than, but consistent with those of the DM$_{IS}$ method within much larger uncertainties, namely
\bea
    \label{eq:f0_BGL}
    f^{BGL}(0) & = &  0.188 \pm 0.074  \qquad  \qquad {\rm  FNAL/MILC+HPQCD +RBC/UKQCD} ~ , ~\nonumber  \\[2mm]
    f^{BGL}(0) & = &  0.218 \pm 0.117  \qquad  \qquad {\rm  HPQCD +RBC/UKQCD}\nonumber ~ . ~
\eea
As already pointed out in Sections\,\ref{sec:FFs} and \ref{sec:unos}, the DM method produces narrower bands for the extrapolation of the FFs with respect to the BGL approach. This is a consequence of the automatic elimination of the subset of input values of the FFs which, although allowed by the uncertainties of the lattice calculations, do not satisfy unitarity (and the KC). See Ref.\,\cite{Simula:2023ujs} for a detailed discussion on the features of the DM unitary filter. 

\begin{table}[htb!]
\renewcommand{\arraystretch}{1.25}
\begin{center}
\begin{tabular}{|c|c|c|c|}
\hline 
\multicolumn{4}{|l|}{\,\,\,\,\,\,\,\,\,\,\,\,\,\,\,\,\,\,\,\,\,\,\,\,\,\,\,\,\,\,\,\,\,\,\,\,\,\,\,\,\,\,\,\,\,\,\,\,\,\,\,\,\,\,\,\,\,\,\,\,\,\,\,\,\,\,\,\,\,\,\,\,\,\,\,\,\,\,\,\,\,\,\,\,\,\,\,\,\,\,\,\,\,\,\,\,\,\,\,\,\,\,\,\, DM$_{IS}$}   \\\hline
 lattice FFs & ~$\vert V_{ub} \vert^{(1)} \cdot 10^{3}$ & ~$\vert V_{ub} \vert^{(2)} \cdot 10^{3}$~ & $\vert V_{ub} \vert^{(3)} \cdot 10^{3}$~\\
\hline 
\cite{Flynn:2023nhi}+\cite{FermilabLattice:2019ikx}+\cite{Bouchard:2014ypa, Monahan:2018lzv}  & ~3.55~(49)~ & ~3.70~(27)~ & ~3.64~(32)~\\
\hline 
\cite{Flynn:2023nhi}+\cite{Bouchard:2014ypa, Monahan:2018lzv}  & ~3.12~(46)~ & ~3.62~(29)~ & ~3.42~(33)~\\
\hline
\multicolumn{4}{|l|}{\,\,\,\,\,\,\,\,\,\,\,\,\,\,\,\,\,\,\,\,\,\,\,\,\,\,\,\,\,\,\,\,\,\,\,\,\,\,\,\,\,\,\,\,\,\,\,\,\,\,\,\,\,\,\,\,\,\,\,\,\,\,\,\,\,\,\,\,\,\,\,\,\,\,\,\,\,\,\,\,\,\,\,\,\,\,\,\,\,\,\,\,\,\,\,\,\,\,\,\,\,\,\,\, BGL}   \\\hline
lattice FFs & ~$\vert V_{ub} \vert^{(1)} \cdot 10^{3}$ & ~$\vert V_{ub} \vert^{(2)} \cdot 10^{3}$~ & $\vert V_{ub} \vert^{(3)} \cdot 10^{3}$~\\
\hline 
\cite{Flynn:2023nhi}+\cite{FermilabLattice:2019ikx}+\cite{Bouchard:2014ypa, Monahan:2018lzv}  & ~4.15~(1.17)~ & ~3.84~(35)~ & ~3.93~(54)~\\
\hline 
\cite{Flynn:2023nhi}+\cite{Bouchard:2014ypa, Monahan:2018lzv}  & ~3.61~(1.43)~ & ~3.74~(44)~ & ~3.69~(76)~\\
\hline
\end{tabular}\\[4mm]
\begin{tabular}{|c|c|c|c|}
\hline
\multicolumn{4}{|l|}{\,\,\,\,\,\,\,\,\,\,\,\,\,\,\,\,\,\,\,\,\,\,\,\,\,\,\,\,\,\,\,\,\,\,\,\,\,\,\,\,\,\,\,\,\,\,\,\,\,\,\,\,\,\,\,\,\,\,\,\,\,\,\,\,\,\,\,\,\,\,\,\,\,\,\,\,\,\,\,\,\,\,\,\,\,\,\,\,\,\,\,\,\,\,\,\,\,\,\,\,\,\,\,\,\,\,\,\,\,\,\,\, DM$_{IS}$}   \\\hline
 lattice FFs & ~$\vert V_{ub} \vert^{(1)} / \vert V_{cb} \vert$ & ~$\vert V_{ub} \vert^{(2)} / \vert V_{cb} \vert$~ & $\vert V_{ub} \vert^{(3)} / \vert V_{cb} \vert$~\\
\hline 
\cite{Flynn:2023nhi}+\cite{FermilabLattice:2019ikx}+\cite{Bouchard:2014ypa, Monahan:2018lzv}  & ~0.085~(13)~ & ~0.088~(8)~ & ~0.087~(9)~\\
\hline 
\cite{Flynn:2023nhi}+\cite{Bouchard:2014ypa, Monahan:2018lzv} & ~0.075~(12)~ & ~0.086~(8)~ & ~0.082~(9)~\\
\hline
\multicolumn{4}{|l|}{\,\,\,\,\,\,\,\,\,\,\,\,\,\,\,\,\,\,\,\,\,\,\,\,\,\,\,\,\,\,\,\,\,\,\,\,\,\,\,\,\,\,\,\,\,\,\,\,\,\,\,\,\,\,\,\,\,\,\,\,\,\,\,\,\,\,\,\,\,\,\,\,\,\,\,\,\,\,\,\,\,\,\,\,\,\,\,\,\,\,\,\,\,\,\,\,\,\,\,\,\,\,\,\,\,\,\,\,\,\,\,\,  BGL}\\\hline
lattice FFs & ~$\vert V_{ub} \vert^{(1)} / \vert V_{cb} \vert$ & ~$\vert V_{ub} \vert^{(2)} / \vert V_{cb} \vert$~ & $\vert V_{ub} \vert^{(3)} / \vert V_{cb} \vert$~\\
\hline 
\cite{Flynn:2023nhi}+\cite{FermilabLattice:2019ikx}+\cite{Bouchard:2014ypa, Monahan:2018lzv}  & ~0.099~(28)~ & ~0.092~(10)~ & ~0.094~(14)~\\
\hline 
\cite{Flynn:2023nhi}+\cite{Bouchard:2014ypa, Monahan:2018lzv} & ~0.086~(34)~ & ~0.089~(11)~ & ~0.088~(19)~\\
\hline
\end{tabular}
\end{center}
\renewcommand{\arraystretch}{1.0}
\caption{\it \small Mean values and uncertainties of the CKM element $\vert V_{ub} \vert$ (top panel) and of the ratio $\vert V_{ub} \vert / \vert V_{cb} \vert$ (bottom panel), obtained using Eqs.\,(\ref{eq:vub}) and (\ref{eq:vubsvcb}) respectively, in the low-$q^2$ region (label $1$), in the high-$q^2$ region (label $2$) or using simultaneously the complete set of experimental data at all $q^2$ (label $3$).  The DM and BGL FFs used to extract these results have been obtained by using either the results of all the three lattice collaborations ({\rm \cite{Flynn:2023nhi}+\cite{FermilabLattice:2019ikx}+\cite{Bouchard:2014ypa, Monahan:2018lzv}}) or only those of RBC/UKQCD and HPQCD ({\rm \cite{Flynn:2023nhi}+\cite{Bouchard:2014ypa, Monahan:2018lzv}}).}
\label{tab:vubavubsvcb}
\end{table}

The main results of our analysis are listed in Table\,\ref{tab:vubavubsvcb}.
As our final result for $\vert V_{ub} \vert$ we quote
\be
    \label{eq:Vub_update}
    \vert V_{ub} \vert =   (3.64  \pm 0.32) \cdot 10^{-3} ~ , ~
\ee
which agrees with the result $\vert V_{ub} \vert  = (3.56 \pm 0.41) \cdot 10^{-3}$ from Ref.\,\cite{Flynn:2023qmi} with a smaller uncertainty. Our finding\,(\ref{eq:Vub_update}) is consistent with the FLAG value\,(\ref{eq:vubflag}), obtained from $B \to \pi \ell  \nu_\ell$ semileptonic decays, within a larger uncertainty as well as with the results of Refs.\,\cite{Biswas:2022yvh, Leljak:2023gna}.
Correspondingly, our best estimate for $\vert V_{ub} \vert / \vert V_{cb} \vert$ is
\be 
    \vert V_{ub} \vert / \vert V_{cb} \vert = 0.087 \pm 0.009 ~ , ~
\ee
which is compatible with the FLAG determination\,\cite{FlavourLatticeAveragingGroupFLAG:2021npn} $\vert V_{ub} \vert / \vert V_{cb} \vert = 0.0844 \pm 0.0056$, used in the UT\emph{fit} analysis of Ref.\,\cite{UTfit:2022hsi} although with a larger error that requires some explanation. 
In Ref.\,\cite{UTfit:2022hsi} the value $\vert V_{ub} \vert / \vert V_{cb} \vert = 0.0844(56)$ was taken from the results quoted in Eqs.\,(311)-(313) of Ref.\,\cite{FlavourLatticeAveragingGroupFLAG:2021npn}, namely
\bea  
    \vert V_{ub} \vert^{(1)} / \vert V_{cb} \vert & = & 0.0819 \pm 0.0072_{lat} \pm 0.0029_{exp}  = 0.0819 \pm 0.0078 \, , \nonumber \\[2mm]
    \vert V_{ub} \vert^{(2)} / \vert V_{cb} \vert & = & 0.0860 \pm 0.0037_{lat} \pm 0.0038_{exp}  = 0.0860 \pm 0.0053 \, ,  \nonumber \\[2mm]
    \vert V_{ub} \vert^{(3)} / \vert V_{cb} \vert & = & 0.0844 \pm 0.0048_{lat} \pm 0.0028_{exp}  = 0.0844 \pm 0.0056 \, . \nonumber  
\eea
We guess that the origin of the difference with the FLAG result in the evaluation of the uncertainty on $\vert V_{ub} \vert / \vert V_{cb} \vert$ is that the fitting procedure of the vector and scalar FFs made by FLAG used data points at $q^2 = 15$ GeV$^2$ (see Fig.\,32 at page 167 of Ref.\,\cite{FlavourLatticeAveragingGroupFLAG:2021npn}), which were never directly computed on the lattice by FNAL/MILC, but instead were obtained indirectly from a Bourrely-Caprini-Lellouch (BCL) fit\,\cite{Bourrely:2008za}. On the contrary, we use values of the FFs computed only in the $q^2$-region covered by direct lattice results (i.e.~$q^2 \gtrsim 17$ GeV$^2$), as shown in Fig.\,\ref{fig:FFBsK}.

\section{Conclusions}
\label{sec:conclusions}

In this work we have applied the DM method of Refs.\,\cite{DiCarlo:2021dzg, Martinelli:2021frl, Martinelli:2021onb, Martinelli:2021myh, Martinelli:2022tte, Martinelli:2022xir} and its new version including the unitary sampling procedure of Ref.\,\cite{Simula:2023ujs}, called DM$_{IS}$, to the determination of the momentum dependence of the FFs entering the semileptonic $B \to D^* \ell \nu_\ell$ decays using the most recent lattice calculations by the FNAL/MILC Collaboration\,\cite{FermilabLattice:2021cdg}, the HPQCD Collaboration\,\cite{Harrison:2023dzh} and the JLQCD Collaboration\,\cite{Aoki:2023qpa}. 
The DM and DM$_{IS}$ results are model-independent and satisfy exactly all the unitary and kinematical constraints. 

We stress that in order to use the $\tau / \mu$ ratio $R(D^*)$ as a test of the Standard model, one has to compare its experimental value with the  theoretical one, obtained exclusively from theoretical calculations of the FFs and not, as done in the past, by fitting the FFs using simultaneously theoretical calculations and experimental data. 
Our  theoretical estimate of $R(D^*)$ is $R(D^*) = 0.262 \pm 0.009$, which differs only by $\simeq 1.5\,\sigma$ from the latest experimental world average $R(D^*)\vert_{\rm{exp}} = 0.284 \pm 0.012$ from HFLAV~\cite{HFLAV_web23}.
We have also computed the theoretical value of other polarization observables, like the longitudinal $D^*$-polarization fraction for heavy and light charged leptons, $F_{L, \tau}$ and $F_{L, \ell}$ ($\ell = e, \mu$) respectively, and the forward-backward asymmetry $A_{FB, \ell}$, obtaining results compatible with available experimental data within $\approx 1$ standard deviation, except for a $\simeq 2.4 \sigma$ difference in the case of $F_{L, \ell}$.

Using the new experimental results for the semileptonic $B \to D^* $ decays by the Belle\,\cite{Belle:2023bwv} and Belle-II\,\cite{Belle-II:2023okj} Collaborations, together with the previous Belle ones from Ref.\,\cite{Belle:2018ezy}, we have also obtained the updated value $\vert V_{cb} \vert = (39.92 \pm 0.64) \cdot 10^{-3}$, which is compatible respectively at the $\simeq 2.6\,\sigma$ and $\simeq 2.0\,\sigma$ level with the most recent inclusive determinations $\vert V_{cb} \vert^{\rm{incl}} = (41.97 \pm 0.48) \cdot 10^{-3}$~\cite{Finauri:2023kte} and $\vert V_{cb} \vert^{\rm{incl}} = (41.69 \pm 0.63) \cdot 10^{-3}$~\cite{Bernlochner:2022ucr}. 
We stress that, since both the exclusive and inclusive determinations of $\vert V_{cb} \vert$ are reaching the percent level of accuracy, it is timely to assess QED effects in $b \to c$ decays form first principles (see also Ref.\,\cite{Bigi:2023cbv}).

From the lattice FFs relevant in semi-leptonic $B_s \to  K$ decays available from Refs.\,\cite{Flynn:2023nhi, FermilabLattice:2019ikx, Bouchard:2014ypa, Monahan:2018lzv} and using the LHCb measurement of ${\cal{BR}}(B_s \to K \ell \nu) /$ ${\cal{BR}}(B_s \to D_s \ell \nu)$ we have also updated the important CKM ratio $\vert V_{ub} \vert/\vert V_{cb} \vert = 0.087 \pm 0.009$. Our result, which is properly based on values of the FFs computed only in the $q^2$-region covered by direct lattice results, is in good agreement with the result $\vert V_{ub} \vert / \vert V_{cb} \vert = 0.0844\pm 0.0056$ from the latest FLAG review\,\cite{FlavourLatticeAveragingGroupFLAG:2021npn}.

\section*{Acknowledgements}
We thank Marco Fedele, Vittorio Lubicz and Luca Silvestrini for many useful discussions. S.S.~is supported by the Italian Ministry of Research (MIUR) under grant PRIN 20172LNEEZ. The work of L.V.~is supported by Agence Nationale de la Recherche (ANR) under contract n. 202650 (ANR-19-CE31-0016, GammaRare).

\appendix

\section{The unitary sampling procedure in the DM method}
\label{sec:appA}
 
In this Appendix, following Ref.\,\cite{Simula:2023ujs}, we describe the main features of the IS procedure applied to the DM method, which allows to generate events for any generic FF satisfying the appropriate unitary constraint for any number of the input data points.
We start by briefly recalling the main features of the DM method applied to the description of a generic FF $f(q^2)$ with definite spin-parity.

\subsection{The DM method}
\label{sec:DM}

Let us consider a set of $N$ values of the FF, $\{ f \} = \{ f(z_j) \}$ with $j = 1, 2, ..., N$, where $z$ is the conformal variable
\be
    \label{eq:z}
    z(q^2) \equiv \frac{\sqrt{t_+ - q^2} - \sqrt{t_+ - t_-}}{\sqrt{t_+ - q^2} + \sqrt{t_+ - t_-}} ~ 
\ee
with $t_\pm \equiv (m_{B} \pm m_{D^*})^2$ in the case of our interest and $z_j \equiv z(q_j^2)$.
Then, the FF at a generic value of $z = z(q^2)$ is bounded by unitarity, analyticity and crossing symmetry to be in the range\,\cite{DiCarlo:2021dzg}
\be
  \beta(z) - \sqrt{\gamma(z)} \leq f(z) \leq \beta(z) + \sqrt{\gamma(z)} ~ , ~
    \label{eq:bounds}
\ee 
where 
\bea
      \label{eq:beta_final}
      \beta(z) & \equiv & \frac{1}{\phi(z, q_0^2) d(z)} \sum_{j = 1}^N f(z_j) \phi(z_j, q_0^2) d_j \frac{1 - z_j^2}{z - z_j} ~ , ~ \\[2mm]
      \label{eq:gamma_final}
      \gamma(z) & \equiv &  \frac{1}{1 - z^2} \frac{1}{\phi^2(z, q_0^2) d^2(z)} \left[ \chi(q_0^2) - \chi_{\{f\}}^{DM}(q_0^2) \right] ~ , ~ \\[2mm]
      \label{eq:chiDM}
      \chi_{\{f\}}^{DM}(q_0^2) & \equiv & \sum_{i, j = 1}^N f(z_i)f(z_j) \phi(z_i, q_0^2)  \phi(z_j, q_0^2) d_i d_j \frac{(1 - z_i^2) (1 - z_j^2)}{1 - z_i z_j} ~ 
\eea
with
\be
    \label{eq:dcoef}
    d(z) \equiv \prod_{m = 1}^N \frac{1 - z z_m}{z - z_m} ~ , ~ \qquad d_j  \equiv \prod_{m \neq j = 1}^N \frac{1 - z_j z_m}{z_j - z_m} ~ . ~
\ee
In Eq.\,(\ref{eq:gamma_final}) the quantity $\chi(q_0^2)$ is the dispersive bound, evaluated at an auxiliary value $q_0^2$ of the squared 4-momentum transfer using suitable two-point correlators\,\cite{Martinelli:2021frl}, and $\phi(z, q_0^2)$ is a kinematical function appropriate for the given form factor\,\cite{Boyd:1997kz}. The kinematical function $\phi$ may contain the contribution of the resonances below the pair production threshold $t_+$.
Following our previous work we adopt the value $q_0^2 = 0$ also in this work. See Section IX of Ref.\,\cite{Simula:2023ujs} for a discussion about the impact of different choices of the value of $q_0^2$.

When $z \to z_i$ one has $d(z) \propto 1 / (z - z_i)$ and, therefore, $\beta(z) \to f_i$ and $\gamma (z) \to 0$. In other words, Eq.\,(\ref{eq:bounds}) exactly reproduces the set of input data $\{ f_i \}$. In a frequentist language this corresponds to a vanishing value of the $\chi^2$-variable.

Unitarity is satisfied only when $\gamma(z) \geq 0$, which implies
\be
    \label{eq:DMfilter}
    \chi(q_0^2) \geq \chi_{\{f\}}^{DM}(q_0^2) ~ . ~
\ee
Such a condition depends on the set of input data $\{ f_i \}$ and it is  independent on any parameterization or fitting Ansatz of the input data. 

The meaning of the DM {\em filter}~(\ref{eq:DMfilter}) is clearer in terms of explicit $z$-expansions, like the BGL ones~\cite{Boyd:1997kz}.
When $\chi \geq \chi_{\{f\}}^{DM}$, it is guaranteed the existence of (at least) one BGL fit (either truncated or untruncated) that satisfies unitarity and, at the same time, reproduces exactly the input data.
On the contrary, when $\chi < \chi_{\{f\}}^{DM}$, a unitary $z$-expansion passing through the data does not exist, since the {\em input data} do not satisfy unitarity.
The important feature of the DM approach is that only the {\it unitary} input data are eligible for consideration, while those data that do not satisfy the unitary filter $\chi \geq \chi_{\{f\}}^{DM}$ are discarded. 

Let us consider a sample of input data points generated through a multivariate Gaussian distributions corresponding to given mean values and covariance matrix.
For each event we can apply the DM filter~(\ref{eq:DMfilter}) and, consequently, we can divide the original sample into two disjoint subsets: the one corresponding to input data satisfying the DM filter and the one made of non-unitary events. Such a separation is not guaranteed by approaches based on explicit $z$-expansions (including the recent Bayesian approach of Ref.\,\cite{Flynn:2023qmi}). Indeed, in these approaches the attention is focused only on the fitting function and not also on the fitted data (either experimental or theoretical ones). Even if the fitting function is constructed to satisfy unitarity, the fitting procedure is applied to all the input data regardless whether they satisfy unitarity or not (i.e., regardless whether the input data can be exactly reproduced by a unitary $z$-expansion).
In the case of the unitary subset of input data it is always possible to find a suitable BGL fit, that satisfies unitarity and at the same time exactly reproduces the input data. This corresponds to the possibility to reach a null value of the $\chi^2$-variable by increasing the order of the truncation of the BGL fit (up to the number of data points).
On the contrary, when the input data do not satisfy the unitary filter, it is not possible to find a fitting $z$-expansion that satisfies unitarity and at the same time exactly reproduces the input data. This corresponds to a non-vanishing value of the $\chi^2$-variable, which depends on the impact of the non-unitary input data. 
In what follows we will refer to the first subset as the {\it unitary} input data and to the second one as the {\it non-unitary} input data.
It is clear that the application of a fitting function (even if unitary) to a subset of input data that do not satisfy unitarity may lead to a distortion of the fitting results related directly to the impact of the non-unitary effects present in the input data. In particular, such a distortion may be relevant when the fitting function extrapolates the form factor in a kinematical region not covered by the input data. 

As shown in Ref.\,\cite{Simula:2023ujs}, the DM filter~(\ref{eq:DMfilter}) may become extremely selective when the number of input data points $N$ increases. 
The origin of this effect is due to the values of the kinematical coefficients $d_j$, given by Eq.~(\ref{eq:dcoef}). These coefficients depend only on the series of values $z_j$ and their numerical values can be quite large in absolute value with alternating signs.
It is therefore very unlikely to generate an event for the form factor points leading to a value of $\chi_{\{f\}}^{DM}(q_0^2)$ as small as $\chi_(q_0^2)$.
A very delicate compensation among the contributions of the various data points to Eq.~(\ref{eq:chiDM}) is required and this naturally implies specific correlations among the form factor points.
In principle, one may increase the size of the sample until some of the events satisfy the unitary filter, but a brute-force increase of the size of the sample may become impracticable for large values of the number of data points $N$.

\subsection{Unitary sampling procedure}
\label{sec:sampling}

The Gaussian multivariate distribution of the input data is based on the probability density function (PDF) given by 
\be
    PDF(f_i) \propto \mbox{exp}\left[ - \frac{1}{2} \sum_{i,j=1}^N (f_i - F_i) C_{ij}^{-1} (f_j - F_j) \right] ~ , ~
    \label{eq:PDF}
\ee
where $\{ F_i \}$ and $\{ C_{ij} \}$ are respectively the mean values and the covariance matrix used as inputs.
As well known, the PDF~(\ref{eq:PDF}) favors the relative likelihood of small values of the quadratic form $\sum_{i,j=1}^N (f_i - F_i) C_{ij}^{-1} (f_j - F_j)$, which however may correspond to large values of the susceptibility\,(\ref{eq:chiDM}).

We now modify the above PDF in order to allow the susceptibility\,(\ref{eq:chiDM}) to be small enough to fulfill the unitary constraint\,(\ref{eq:DMfilter}).
We consider the following new PDF:
\bea
    \label{eq:PDF_DM}
    PDF_{IS}(f_i) & \propto &  PDF(f_i) \cdot \mbox{exp}\left[ - \frac{s}{\chi(q_0^2)} \chi_{\{f\}}^{DM}(q_0^2) \right] ~ \\[2mm]
                          & \propto & \mbox{exp}\left[ - \frac{1}{2} \sum_{i,j=1}^N (f_i - F_i) C_{ij}^{-1} (f_j - F_j) - \frac{s}{\chi(q_0^2)} 
                                             \sum_{i,j=1}^N f_i D_{ij}^{-1}(q_0^2) f_j \right] ~ , ~ \nonumber
\eea
where $s$ is a parameter and the matrix $D^{-1}(q_0^2)$ is defined as
\be
    D_{ij}^{-1}(q_0^2) \equiv \frac{\phi(z_i, q_0^2) d_i (1 - z_i^2) ~ \phi(z_j, q_0^2) d_j (1 - z_j^2)}{1 - z_i z_j} ~ .
    \label{eq:chiDM_ij}
\ee
The use of Eq.\,(\ref{eq:PDF_DM}) as a PDF allows to increase the relative likelihood of small values of the susceptibility $\chi_{\{f\}}^{DM}(q_0^2)$ at the expense of decreasing the PDF\,(\ref{eq:PDF}).
Introducing the matrix $\widetilde{C}$  defined in compact notation as
\be
    \label{eq:Ctilde}
    \widetilde{C}^{-1} = C^{-1} + \frac{2s}{\chi(q_0^2)} D^{-1}(q_0^2) ~ , ~
\ee
Eq.\,(\ref{eq:PDF_DM}) can be easily rewritten in the form
\be
     PDF_{IS}(f_i) \propto \mbox{exp}\left[ - \frac{1}{2} \sum_{i,j=1}^N (f_i - \widetilde{F}_i) \widetilde{C}_{ij}^{-1} (f_j - \widetilde{F}_j) ~ 
                                       - \frac{1}{2} \sum_{i,j=1}^N (F_i - \widetilde{F}_i) C_{ij}^{-1} F_j \right] ~ , ~
    \label{eq:PDF_DM_alt}
\ee
where the new vector of mean values $\widetilde{F}$ is related to the starting one $F$ by
\be
    \label{eq:Ftilde}
    \widetilde{F} = \widetilde{C} ~ C^{-1} ~ F ~ . ~
\ee
Note that the second exponential in the r.h.s.~of Eq.\,(\ref{eq:PDF_DM_alt}) does not depend on $\{ f_i \}$ and therefore it is irrelevant for the relative likelihood of the events, so that the new PDF is simply given by
\be
     PDF_{IS}(f_i) \propto \mbox{exp}\left[ - \frac{1}{2} \sum_{i,j=1}^N (f_i - \widetilde{F}_i) \widetilde{C}_{ij}^{-1} (f_j - \widetilde{F}_j) \right]~ , ~
    \label{eq:PDF_DM_final}
\ee
which represents a multivariate gaussian distribution characterized by the new set of input values $\{ \widetilde{F}_i \}$ and $\{ \widetilde{C}_{ij} \}$ given by Eqs.\,(\ref{eq:Ftilde}) and (\ref{eq:Ctilde}), respectively.

The parameter $s$ in Eq.\,(\ref{eq:Ctilde}) governs the number of events satisfying the unitary filter~(\ref{eq:DMfilter}). The latter one increases as $s$ increases and both the new mean values $\widetilde{F}$ and the new covariance matrix $\widetilde{C}$ depend on the value of the parameter $s$. In order to get rid off such a dependence, one can adopt a simple iterative procedure, in which the DM filter is applied again to the new set of input values and covariance matrix until convergence is reached. This procedure is illustrated in detail in Ref.\,\cite{Simula:2023ujs} in the case of the electromagnetic pion forrn factor.

In the case of the the $B \to D^* \ell \nu_\ell$ decays there are three unitarity constraints on the FFs:
\bea
   \label{eq:DMfilter1}
    \chi_{1^-}(q_0^2) & \geq & \chi_{\{ g \}}^{DM}(q_0^2) ~ , ~ \\[2mm]
   \label{eq:DMfilter2}
    \chi_{1^+}(q_0^2) & \geq & \chi_{\{ f \}}^{DM}(q_0^2) + \chi_{\{ F_1 \}}^{DM}(q_0^2) ~ , ~ \\[2mm]
   \label{eq:DMfilter3}
    \chi_{0^-}(q_0^2) & \geq & \chi_{\{ F_2 \}}^{DM}(q_0^2) ~ , ~
\eea
where $\chi_{\{ g, f,  F_1, F_2 \}}^{DM}(q_0^2)$ are given by Eq.\,(\ref{eq:DMfilter}) in terms of the known values of the corresponding FF and appropriate kinematical function $\phi$\,\cite{Boyd:1997kz}, namely for $q_0^2 = 0$
\bea
    \label{eq:phig}
    \phi_{g}(z, 0) & = & 16 r^2 \sqrt{\frac{2}{3\pi}} \frac{(1 + z)^{2}}{\sqrt{1 - z}\left[ (1 + r)(1 - z) + 2 \sqrt{r}(1 + z) \right]^4} ~ , ~ \nonumber \\[2mm]
    \label{eq:phif}
    \phi_{f}(z, 0) & = & 4 \frac{r}{m_{B}^2} \sqrt{\frac{2}{3\pi}} \, \frac{(1 + z)(1 - z)^{3/2}}{\left[ (1 + r)(1 - z) + 2 \sqrt{r}(1 + z) \right]^4} ~ , ~ \nonumber \\[2mm ]
    \label{eq:phiF1}
    \phi_{F_1}(z, 0) & = & 4 \frac{r}{m_{B}^3} \sqrt{\frac{1}{3\pi}} \frac{(1 + z)(1 - z)^{5/2}}{\left[ (1 + r)(1 - z) + 2 \sqrt{r}(1 + z) \right]^5} ~ , ~ \\[2mm]
    \label{eq:phiF2}
    \phi_{F_2}(z, 0) & = &  16 \, r (1 + r)^2 \sqrt{\frac{1}{\pi}} \frac{(1 + z)^{2}}{\sqrt{1 - z}\left[ (1 + r)(1 - z) + 2 \sqrt{r}(1 + z) \right]^4} ~ \nonumber 
\eea
with $r \equiv m_{D^*} / m_B$.
The presence of resonances below the pair production threshold lead to the following modification of the kinematical function $\phi(z, q_0^2)$
\be
    \label{eq:poles}
    \phi(z, q_0^2) \to \phi(z, q_0^2) \cdot \prod_R \frac{z - z(m_R^2)}{1 - z \, z(m_R^2)} ~ , ~
\ee
where $m_R$ is the mass of the resonance $R$. For the masses of the poles corresponding to $B_c^{(*)}$ mesons with different quantum numbers entering the various FFs we refer to Table III of Ref.\,\cite{Bigi:2017jbd}.

In Eqs.\,(\ref{eq:DMfilter1})-(\ref{eq:DMfilter3}) the susceptibilities $\chi_{1^-}(q_0^2) $, $\chi_{1^+}(q_0^2) $ and $\chi_{0^-}(q_0^2)$ represent the dispersive bounds for the corresponding channels with definite spin-parity. They have been computed nonperturbatively on the lattice in Ref.\,\cite{Martinelli:2021frl} at $q_0^2 = 0$, namely
\bea
      \label{eq:bound1-}
     \chi_{1^-}(0) & = & (5.84 \pm 0.44) \cdot 10^{-4} ~ \mbox{GeV}^{-2} ~ , ~ \nonumber \\[2mm]
     \label{eq:bound1+}
     \chi_{1^+}(0) & = & (4.69 \pm 0.30) \cdot 10^{-4} ~ \mbox{GeV}^{-2} ~ , ~ \nonumber \\[2mm]
     \label{eq:bound0-}
     \chi_{0^-}(0) & = & (21.9 \pm 1.9) \cdot 10^{-3} ~ . ~\nonumber     
\eea

Thus, we need to generalize Eq.\,(\ref{eq:Ctilde}) to include the contributions related to the various FFs.  This can be easily done and leads to the following  inverse of the new covariance matrix $\widetilde{C}$:
\be
    \label{eq:Ctilde_BDstar}
    \widetilde{C}^{-1} = C^{-1} + \frac{2s_g}{\chi_{1^-}(q_0^2)} D_g^{-1}(q_0^2) + \frac{2s_f}{\chi_{1^+}(q_0^2)} D_f^{-1}(q_0^2) + 
    \frac{2s_{F_1}}{\chi_{1^+}(q_0^2)} D_{F_1}^{-1}(q_0^2) + \frac{2s_{F_2}}{\chi_{0^-}(q_0^2)} D_{F_2}^{-1}(q_0^2) ~ , ~
\ee
where by construction we impose $s_f = s_{F_1}$\footnote{In the numerical applications we have always assumed $s_g = s_f = s_{F_1} = s_{F_2}$.} and the matrices $D_{g, f, F_1, F_2}^{-1}$ are simple generalizations of Eq.\,(\ref{eq:chiDM_ij}) using the appropriate kinematical function $\phi_{g, f, F_1, F_2}$.

\subsection{Implementation of the kinematical constraints}
\label{sec:appC}

The implementation of the KCs\,(\ref{eq:KCs}) within the DM method is straightforward and it is described in detail in the Appendix C of Ref.\,\cite{Martinelli:2021myh}. 
Here, we limit ourselves to observe that the application of the KCs can be quite selective as in the case of the unitary filters and it may lead to quite few surviving events from the initial sample.
In this respect we note that the IS procedure, described in Section\,\ref{sec:sampling}, can be easily adapted also to take care of the KCs.

Let us consider, for instance, the KC at $w = 1$, i.e.~$z_{KC} = 0$. Using the central values of the FFs $f(1)$ and $F_1(1)$ predicted by the DM method in Eq.\,(\ref{eq:beta_final}), the KC can be rewritten in the form
\be
     \sum_{j = 1}^N \Phi_{KC}(z_j, q_0^2) f(z_j) = 0 ~ , ~
\ee
where the vector $\Phi_{KC}(z_j, q_0^2)$ is given by
\be
      \Phi_{KC}(z_j, q_0^2) = m_B (1 - r) \frac{d_j}{d(z_{KC})} \frac{1 - z_j^2}{z_{KC} - z_j} \frac{\phi_f(z_j, q_0^2)}{\phi_f(z_{KC}, q_0^2)} ~
\ee
when $j$ refers to the FF $f$,
\be
      \Phi_{KC}(z_j, q_0^2) = -  \frac{d_j}{d(z_{KC})} \frac{1 - z_j^2}{z_{KC} - z_j} \frac{\phi_{F_1}(z_j, q_0^2)}{\phi_{F_1}(z_{KC}, q_0^2)} ~
\ee 
when $j$ refers to the FF $F_1$ and zero otherwise.
Thus, we can add to the PDF\,(\ref{eq:PDF_DM_final}) a contribution quadratic in the FFs, which leads to the following addition to the matrix matrix $\widetilde{C}^{-1}$:
\be
     \widetilde{C}_{ij}^{-1} \to \widetilde{C}_{ij}^{-1} + s_{KC} \, \Phi_{KC}(z_i, q_0^2) \, \Phi_{KC}(z_j, q_0^2) 
\ee
where $s_{KC}$ is a parameter, whose value can be chosen to increase the number of events fulfilling the given KC.

\section{Covariance matrices for various theoretical observables}
\label{sec:appD}

In this Appendix we provide the full covariance matrices of the set of pure theoretical observables given in Eq.\,(\ref{eq:set_cov}), namely $ \{ \Gamma_\tau / |V_{cb}|^2, \Gamma_\ell / |V_{cb}|^2, R(D^*), P_\tau(D^*), F_{L, \tau}, F_{L, \ell}, A_{FB, \ell} \} $, corresponding to the lattice FFs obtained via the DM method applied separately to each of the three lattice calculations of Refs.\,\cite{FermilabLattice:2021cdg, Harrison:2023dzh, Aoki:2023qpa} and to all the lattice results from the three Collaborations.
Expressing $\Gamma_\tau / |V_{cb}|^2$ and $\Gamma_\ell / |V_{cb}|^2$ in units of $10^{11}$ GeV we get:

\begin{itemize}

\item{FNAL/MILC\,\cite{FermilabLattice:2021cdg}}
\bea
&&\mbox{mean values: } \{0.3198, 1.1636, 0.2753, -0.5293, 0.4177, 0.4501, 0.2614\}, \nonumber \\[2mm]
&&\mbox{errors: } \{0.0138, 0.0771, 0.0082, 0.0072, 0.0088, 0.0189, 0.0137\}, \nonumber \\[2mm]
&&\mbox{correlations: } \left( 
\begin{tabular}{ccccccc}
     1.000&     0.942&    -0.656&     0.678&     0.577&     0.655&    -0.335 \\[2mm]
     0.942&     1.000&    -0.870&     0.760&     0.585&     0.701&    -0.313 \\[2mm]
    -0.656&    -0.870&     1.000&    -0.718&   -0.472&    -0.620&     0.212 \\[2mm]
     0.678&     0.760&    -0.718&     1.000&     0.840&     0.845&    -0.389 \\[2mm]
     0.577&     0.585&    -0.472&     0.840&     1.000&     0.955&    -0.678 \\[2mm]
     0.655&     0.701&    -0.620&     0.845&     0.955&     1.000&    -0.683 \\[2mm]
    -0.335&    -0.313&     0.212&    -0.389&    -0.678&    -0.683&    1.000
\end{tabular}
\right). \nonumber
\eea

\item{HPQCD\,\cite{Harrison:2023dzh}}
\bea
&&\mbox{mean values: } \{0.3214, 1.2137, 0.2657, -0.5426, 0.3990, 0.4347, 0.2652\}, \nonumber \\[2mm]
&&\mbox{errors: } \{0.0170, 0.1079, 0.0115, 0.0181, 0.0232, 0.0418, 0.0299\}, \nonumber \\[2mm]
&&\mbox{correlations: } \left( 
\begin{tabular}{ccccccc}
     1.000&     0.937&    -0.703&       0.547&     0.503&     0.544&     -0.352 \\[2mm]
     0.937&     1.000&    -0.904&       0.685&     0.599&     0.698&     -0.442 \\[2mm]
    -0.703&    -0.904&     1.000&      -0.736&    -0.611&   -0.767&      0.476 \\[2mm]
     0.547&     0.685&    -0.736&      1.000 &     0.918&     0.925&     -0.612 \\[2mm]
     0.503&     0.599&    -0.611&       0.918&     1.000&     0.951&     -0.686 \\[2mm]
     0.544&     0.698&    -0.767&       0.925&     0.951&     1.000&     -0.731 \\[2mm]
    -0.352&   -0.442&     0.476&      -0.612&    -0.686&    -0.731&      1.000
\end{tabular}
\right). \nonumber
\eea

\item{JLQCD\,\cite{Aoki:2023qpa}}
\bea
&&\mbox{mean values: } \{0.3330, 1.3490, 0.2474, -0.5086, 0.4478, 0.5160, 0.2201\}, \nonumber \\[2mm]
&&\mbox{errors: } \{0.0156, 0.0991, 0.0076, 0.0109, 0.0161, 0.0287, 0.0210\}, \nonumber \\[2mm]
&&\mbox{correlations: } \left( 
\begin{tabular}{ccccccc}
     1.000&     0.965&    -0.780&     0.570&     0.540&     0.556&    -0.243 \\[2mm]
     0.965&     1.000&    -0.915&     0.687&     0.661&     0.685&    -0.346 \\[2mm]
    -0.780&    -0.915&     1.000&    -0.771&    -0.755&    -0.790&     0.456 \\[2mm]
     0.570&     0.687&    -0.771&     1.000&     0.867&     0.874&    -0.563 \\[2mm]
     0.540&     0.661&    -0.755&     0.867&     1.000&     0.985&    -0.714 \\[2mm]
     0.556&     0.685&    -0.790&     0.874&     0.985&     1.000&    -0.701 \\[2mm]
    -0.243&    -0.346&     0.456&    -0.563&    -0.714&    -0.701&     1.000
\end{tabular}
\right). \nonumber
\eea

\item{FNAL/MILC+HPQCD+JLQCD\,\cite{FermilabLattice:2021cdg, Harrison:2023dzh, Aoki:2023qpa}}
\bea
&&\mbox{mean values: } \{0.3296, 1.2752, 0.258694, -0.5213, 0.4250, 0.4733, 0.2518\}, \nonumber \\[2mm]
&&\mbox{errors: } \{0.0089, 0.0537, 0.0047, 0.0057, 0.0074, 0.0144, 0.0101\}, \nonumber \\[2mm]
&&\mbox{correlations: } \left( 
\begin{tabular}{ccccccc}
     1.000&    0.956&   -0.735&    0.631&    0.519&    0.573&   -0.339 \\[2mm]
     0.956&    1.000&   -0.901&    0.732&    0.603&    0.677&   -0.394 \\[2mm]
    -0.735&   -0.901&    1.000&   -0.761&  -0.629&   -0.722&    0.411 \\[2mm]
     0.631&    0.732&   -0.761&    1.000&    0.876&    0.871&   -0.503 \\[2mm]
     0.519&    0.603&   -0.629&    0.876&    1.000&    0.970&   -0.712 \\[2mm]
     0.573&    0.677&   -0.722&    0.871&    0.970&    1.000&   -0.725 \\[2mm]
    -0.339&   -0.394&    0.411&   -0.503&   -0.712&  -0.725&    1.000
\end{tabular}
\right). \nonumber
\eea

\end{itemize}

\bibliography{biblio1}
\bibliographystyle{EPJC}

\end{document}